\documentclass[12pt,preprint]{aastex}

\usepackage{amssymb}
\usepackage{amsmath}
\usepackage{graphicx}
\usepackage[normalem]{ulem}
\usepackage{natbib}
\usepackage[dvips]{color}

\renewcommand\sout{\bgroup \color{red} \ULdepth=-.5ex \ULset}

\shorttitle{Nuclear constraints on properties of neutron star
crusts}\shortauthors{Xu, Chen, Li, Ma}

\begin{document}
\title{Nuclear constraints on properties of neutron star crusts}
\author{Jun Xu$^{1,2}$, Lie-Wen Chen$^{1,3}$, Bao-An Li$^{4}$, Hong-Ru
Ma$^{1}$}

\affil{$^1$Institute of Theoretical Physics, Shanghai Jiao Tong
University, Shanghai 200240, China} \affil{$^2$Cyclotron Institute
and Physics Department, Texas A\&M University, College Station,
Texas 77843-3366, USA} \affil{$^3$Center of Theoretical Nuclear
Physics, National Laboratory of Heavy-Ion Accelerator, Lanzhou,
730000, China} \affil{$^4$Department of Physics, Texas A\&M
University-Commerce, Commerce, Texas 75429-3011, USA}
\email{xujun@comp.tamu.edu, lwchen@sjtu.edu.cn,
Bao-An\_Li@tamu-commerce.edu, hrma@sjtu.edu.cn}

\begin{abstract}
The transition density $\rho _{t}$ and pressure $P_{t}$ at the inner
edge separating the liquid core from the solid crust of neutron
stars are systematically studied using a modified Gogny (MDI) and
$51$ popular Skyrme interactions within well established dynamical
and thermodynamical methods. First of all, it is shown that the
widely used parabolic approximation to the full Equation of State
(EOS) of isospin asymmetric nuclear matter may lead to huge errors
in estimating the transition density and pressure, especially for
stiffer symmetry energy functionals $E_{sym}(\rho)$, compared to
calculations using the full EOS within both the dynamical and
thermodynamical methods mainly because of the energy curvatures
involved. Thus, fine details of the EOS of asymmetric nuclear matter
are important for locating accurately the inner edge of the neutron
star crust. Secondly, the transition density and pressure decrease
roughly linearly with the increasing slope parameter $L$ of the
$E_{sym}(\rho)$ at normal nuclear matter density using the full EOS
within both the dynamical and thermodynamical methods. It is also
shown that the thickness, fractional mass and moment of inertia of
neutron star crust are all very sensitive to the parameter $L$
through the transition density $\rho _{t}$ whether one uses the full
EOS or its parabolic approximation. Moreover, it is shown that the
$E_{sym}(\rho)$ constrained in the same sub-saturation density range
as the neutron star crust by the isospin diffusion data in heavy-ion
collisions at intermediate energies limits the transition density
and pressure to $0.040$ fm$^{-3}$ $\leq \rho _{t}\leq 0.065$
fm$^{-3}$ and $0.01$ MeV/fm$^{3}$ $\leq P_{t}\leq 0.26$
MeV/fm$^{3}$, respectively. These constrained values for the
transition density and pressure are significantly lower than their
fiducial values currently used in the literature. Furthermore, the
mass-radius relation and several other properties closely related to
the neutron star crust are studied by using the MDI interaction. It
is found that the newly constrained $\rho_t$ and $P_t$ together with
the earlier estimate of $\Delta I/I>0.014$ for the crustal fraction
of the moment of inertia of the Vela pulsar impose a more stringent
constraint of $R\geq 4.7+4.0M/M_{\odot}$ km for the radius $R$ and
mass $M$ of neutron stars compared to previous studies in the
literature.
\end{abstract}

\maketitle

\section{Introduction}
\label{introduction}

Neutron stars are among the most mysterious objects in the Universe.
They are natural testing grounds of our knowledge about the Equation
of State (EOS) of neutron-rich nuclear matter. The latter determines
the structure and many properties of neutron
stars~\citep{Lat07,Oya07,Dou00,Hor01,Kub07,Chomaz07,Rab08}. Neutron
stars are expected to have a solid inner crust which is believed to
play an important role in understanding a number of astrophysical
observations~\citep{BPS71,BBP71,Pet95a,Pet95b,Lat00,Ste05,Lat07,Cha08},
such as, pulsar glitches \citep{Lin99}, quasi-periodic oscillations
observed in x-ray emission following x-ray bursts on neutron star
\citep{Dun98}, the cooling observed over the first several years
following superbursts from neutron stars or giant flares from
magnetars \citep{Rut06}, and neutrino opacities \citep{Hor04,Bur06}.
The solid inner crust of a neutron star comprises the region between
the density $\rho _{out}$ where neutrons drip out of nuclei and the
density $\rho _{t}$ where the transition to the homogeneous
nucleonic matter occurs. While the $\rho _{out}$ is relatively well
determined to be $\rho _{out}\approx 4\times 10^{11}$ g/cm$^{3}$
\citep{Rus06,Hem08}, the transition density $\rho _{t}$ is still
very uncertain \citep{Lat00,Lat07}. This is largely due to our poor
knowledge about the EOS of neutron-rich nuclear matter, especially
the density dependence of the nuclear symmetry energy
$E_{sym}(\rho)$ at sub-saturation densities \citep{Lat00,Lat07}.
Consequently, our ability of understanding accurately many important
properties of neutron stars has been
hampered~\citep{Lat04,Lat00,Lat07}.

The EOS of neutron-rich nuclear matter also plays an important role
in heavy-ion collisions especially those induced by neutron-rich
radioactive beams in terrestrial laboratories. While heavy-ion
collisions are not expected to create the same matter and conditions
as in neutron stars, the same elementary nuclear interactions are at
work in the two cases. Thus, it is important to examine
ramifications of conclusions regrading the EOS extracted from one
field in the other one. Significant progress has been made recently
in constraining the EOS of neutron-rich nuclear matter using
heavy-ion experiments (See, e.g., ref.~\citep{LCK08} for the most
recent review). In particular, compared to the existing model
predictions in the literature the analyses of isospin diffusion
data~\citep{Tsa04,Che05a,LiBA05c} in heavy-ion collisions have
constrained relatively tightly the $E_{sym}(\rho)$ in exactly the
same sub-saturation density region around the expected inner edge of
neutron star crust. Moreover, conclusions from analyzing some recent
data~\citep{She07} of the isoscaling phenomenon~\citep{Tsa01} in
heavy-ion collisions and the available data on the thickness of
neutron-skin in $^{208}$Pb \citep{Ste05b,LiBA05c,Che05b} are
consistent with the $E_{sym}(\rho)$ constrained by the isospin
diffusion data. Furthermore, the lower bound of the experimentally
constrained $E_{sym}(\rho)$ is consistent with the Relativistic Mean
Field model prediction using the FSUGold interaction that can
reproduce not only saturation properties of nuclear matter but also
structure properties and giant resonances of many finite
nuclei~\citep{Piek07}. While some model dependence and uncertainties
still exist in the analyses of the above mentioned experiments and
calculations, an overlapping area of the extracted $E_{sym}(\rho)$
from several analyses has appeared in the sub-saturation density
region~\citep{Tsa08,Lyn09}. On the other hand, extremely impressive
progress has also been made in astrophysical observations relevant
for constraining the EOS of nuclear matter. To our best knowledge,
nevertheless, mainly because of the low precision associated with
the current measurements of neutron star radii, a non-controversial
conclusion on the EOS and the density dependence of symmetry energy
has yet to come. More accurate observations of neutron stars
properties, especially their radii, with advanced x-ray satellites
and other observatories, will hopefully enable us to constrain
stringently the EOS of neutron-rich matter in the near future. A
direct cross-check on the EOS extracted independently from heavy-ion
reactions and neutron star observations will then be possible. In
the meantime, examinations of astrophysical implications of the EOS
constrained by heavy-ion reactions are useful. At the WCI3 meeting
in 2005, Horowitz suggested the heavy-ion physics community to
investigate whether one can use the information from heavy-ion
collisions to constrain the core-crust transition density in neutron
stars~\citep{Hor05}. It is thus interesting to investigate timely
how the behaviors of the $E_{sym}(\rho)$ constrained at
sub-saturation densities by heavy-ion experiments may help limit the
transition density $\rho _{t}$ and pressure $P_{t}$ at the inner
edge of neutron stars~\citep{Xu09}.

To our best knowledge, all existing studies indicate consistently
that the transition density is very sensitive to the density
dependence of the nuclear symmetry
energy~\citep{Lat07,Oya07,Dou00,Kub07}. Very often, the so-called
parabolic approximation (PA) to the EOS of isospin asymmetric
nuclear matter is used. While the PA is mathematically valid only at
small isospin asymmetries, interestingly, it has been found
empirically true even for large isospin asymmetries for nucleonic
mater using most models and interactions, see, e.g.,
refs.~\citep{Bom91,Che01,Zuo03,Xu07a,Dal07,Mou07}. Nevertheless,
since the $npe$ matter in the crust at $\beta$-equilibrium is highly
neutron rich and the determination of the transition density depends
on the second order derivatives of the energy density, the fine
details of the EOS can influence the transition density
significantly as first pointed out by Arponen in 1972~\citep{Arp72}.
It is thus interesting and necessary to compare calculations using
both the full EOS and its parabolic approximation. Indeed, we found
that the PA leads to significantly different transition density and
pressure compared to the calculations using the full EOS. It should
be mentioned that the PA may also significantly modify the proton
fraction in $\beta$-equilibrium neutron-star matter and the critical
density for the direct Urca process which can lead to faster cooling
of neutron stars~\citep{Zha01,Ste06}. To investigate effects of
nuclear interactions we use a modified Gogny (MDI) and $51$ Skyrme
interactions widely used in the literature. The same MDI interaction
has been used in extracting the $E_{sym}(\rho)$ from heavy-ion
reactions within a transport model~\citep{Che05a,LiBA05c}. Using the
$E_{sym}(\rho)$ constrained by the isospin diffusion
data~\citep{Tsa04}, we can put a constraint on the transition
density and pressure, respectively. We will then examine the
implications of these constraints on the mass-radius correlation and
the crustal fraction of the moment of inertia of neutron stars.

This paper is organized as follows. In Section \ref{two methods} we
briefly review the dynamical and thermodynamical methods widely used
for locating the inner edge of neutron star crust, and derive their
relationship analytically. In Section \ref{EOS} we summarize the EOS
and symmetry energy obtained using the MDI interaction and $51$
Skyrme interactions within the Hartree-Fock approach. We also
examine the associated proton fraction and several thermodynamical
properties including the energy density, pressure and the speed of
sound in neutron star matter at $\beta$-equilibrium. The general
formalisms for describing the structure of neutron stars are
outlined in Section \ref{nstarth}. We thus present the results of
our calculations and discuss several important issues regarding the
transition density and the structure of neutron stars in Section
\ref{results}. A summary is given in Section \ref{summary}.

\section{Methods for locating the inner edge of neutron star crust}
\label{two methods}

The inner edge of neutron star crust corresponds to the phase
transition from the homogeneous matter at high densities to the
inhomogeneous matter at low densities. In principle, the transition
density $\rho _{t}$ can be obtained by comparing a detailed model of
the nonuniform solid crust to the uniform liquid core in the neutron
star. While this is practically very difficult since the inner crust
may have a very complex structure, usually known as ``nuclear pasta"
\citep{Rav83,Has84,Lor93,Oya93,Hor04,Ste08,Gog08,Ava08,Ava08b}, it
can be explored within several approaches including the molecular
dynamics simulations~\citep{Wat05,Hor06} and the 3D Hartree-Fock
model~\citep{New09}. Furthermore, the core-crust transition is
thought to be a very weak first-order phase transition and model
calculations lead to very small density discontinuities at the
transition~\citep{Pet95b,Dou00,Dou01,Hor03}. Alternatively, a well
established approach for estimating the $\rho _{t}$ is to search for
the density at which the uniform liquid first becomes unstable
against small-amplitude density fluctuations, indicating the start
of forming nuclear clusters.Although some quantum effects such as
the shell effects in more microscopic methods may influence the
core-crust transition density, this approach has been shown to
produce a very small error for the actual core-crust transition
density and it would yield the exact transition density for a
second-order phase transition~\citep{Pet95b,Dou00,Dou01,Hor03}.
Presently, there are several such methods, such as, the dynamical
method~\citep{BPS71,BBP71,Pet95a,Pet95b,Dou00,Oya07,Chomaz07}, the
thermodynamical method~\citep{Kub07,Lat07,Wor08,Kub07,Lat07} and the
Random Phase Approximation (RPA) \citep{Hor01,Hor03}. In the present
work, we use both the dynamical and thermodynamical methods.

In the following, we will first review briefly the dynamical method
and the thermodynamical method, separately. While they are both well
established and applied extensively in studying not only the
core-crust transition in neutron stars but also the liquid-gas phase
transition in asymmetric nuclear matter, somewhat different results
are often obtained. It is thus necessary to study in detail the
differences and relations between them. We shall first show
analytically that the thermodynamical method corresponds to the
long-wavelength limit of the dynamical method when the Coulomb
interaction is neglected, and then compare numerically their
predictions.

\subsection{The dynamical method}
\label{dynamical}

To describe small density fluctuations in the $npe$ matter, one
can write the density of particle $q \in \{n,p,e\}$
as~\citep{BPS71,Pet95b,Chomaz07}
\begin{equation}
\rho_q = \rho^0_q + \delta \rho_q.
\end{equation}
The density variation can be decoupled into plane-waves
\begin{equation}
\delta \rho_q = A_q e^{i\vec k \cdot \vec r} + c.c.,
\end{equation}
of wave vector $\vec k$ and amplitude $A_q$. This kind of density
variation occurs when a momentum $\vec k$ is transferred to the
particle system, e.g., through collisions and the ``dynamical
method" is named after this. It has been shown that the variation
of the free energy density generated by the density fluctuation of
amplitude $\tilde{A}=\left ( A_n, A_p, A_e\right )$ can be written
as~\citep{BPS71,Pet95b,Chomaz07}
\begin{equation}
\delta f={\tilde{A}^*} \mathcal C^f \tilde{A},
\end{equation}
where
\begin{eqnarray}
\label{dymethod} C^f &=& \left(
\begin{array}{ccc}
\partial\mu _{n}/\partial\rho _{n} & \partial\mu _{n}/\partial\rho _{p} & 0\\
\partial\mu _{p}/\partial\rho _{n} & \partial\mu _{p}/\partial\rho _{p} & 0\\
0 & 0 & \partial\mu _{e}/\partial\rho _{e}\\
\end{array}
\right) \notag \\
&+& k^2 \left(
\begin{array}{ccc}
D_{nn} & D_{np} & 0\\
D_{pn} & D_{pp} & 0\\
0 & 0 & 0\\
\end{array}
\right) + \frac{4\pi e^2}{k^2} \left(
\begin{array}{ccc}
0 & 0 & 0\\
0 & 1 & -1\\
0 & -1 & 1\\
\end{array}
\right)
\end{eqnarray}
is the free-energy curvature matrix. The instability region of the
$npe$ matter can be located by examining when the convexity of the
free-energy curvature matrix is violated.
The convexity of the matrix $C^f$ requires that
\begin{equation}\label{cmatrix}
C^f_{11}>0~\text{or}~C^f_{22}>0, \left|
\begin{array}{ll}
C^f_{11} & C^f_{12}\\
C^f_{21} & C^f_{22}\\
\end{array}
\right|>0, \left|
\begin{array}{ccc}
C^f_{11} & C^f_{12} & C^f_{13}\\
C^f_{21} & C^f_{22} & C^f_{23}\\
C^f_{31} & C^f_{32} & C^f_{33}\\
\end{array}
\right|>0.
\end{equation}
Here $C^f_{33}$ is always positive so we do not take it into
consideration. If the system stays stable, the convexity of the
matrix $C^f$ should be retained for all values of $k$. The first
term in the right hand of Eq.~(\ref{dymethod}) is the bulk term,
which just defines the stability condition of the nuclear matter
part as will be shown later. The second term in the right hand of
Eq.~(\ref{dymethod}) describes the contribution of the density
gradient. For the Skyrme-Hartree-Fock (SHF) model~\citep{Chabanat}
one has
\begin{equation}
D_{nn}=D_{pp} =\frac{3}{16} \left[ t_1(1-x_1) - t_2(1+x_2) \right],
\end{equation}
\begin{equation}
D_{np}=D_{pn} =\frac{1}{16} \left[ 3t_1(2+x_1) - t_2(2+x_2)
\right],
\end{equation}
in terms of the standard Skyrme interaction parameters $x_1, x_2,
t_1$ and $t_2$. The MDI interaction, however, does not have a
gradient term. To remedy this drawback we set
$D_{pp}=D_{nn}=D_{np}=132$ MeV$\cdot $fm$^{5}$ as used in the work
by Oyamatsu et al.~\citep{Oya07} when we apply the MDI interaction.
This choice is quite consistent with the empirical values from the
SHF calculations. We note here that the averaged value of
$D_{pp}=D_{nn}$ and $D_{np}=D_{pn}$ is, respectively, $140.9$ and
$118.8$ MeV$\cdot $fm$^{5}$ for the $51$ Skyrme forces we will use
in the following. Furthermore, as we will show later, the transition
density and pressure are rather insensitive to the variation of
$D_{pp}=D_{nn}$ and $D_{np}=D_{pn}$. The last term in the right hand
of Eq.~(\ref{dymethod}) is the Coulomb term, which is generated by
the Coulomb interactions of electrons and protons. It should be
noted that additional $k$-dependent terms due to the finite range of
the MDI interaction via exchange terms as well as the Coulomb
exchange terms are neglected in Eq.~(\ref{dymethod}). As we will
show later, the bulk term dominates the result and the density
gradient term and Coulomb term are not important for the
determination of the transition density and the associated
transition pressure. The density gradient term and Coulomb term
usually make the system slightly more stable and thus reduce
correspondingly the region of instability.

For small density fluctuations, to guarantee the convexity of the
curvature matrix it is sufficient for the last determinant in
Eq.~(\ref{cmatrix}) to be positive~\citep{BBP71,Pet95b}, i.e.,
\begin{equation}\label{Vdyn}
V_{dyn}(k) \approx V_0 + \beta k^2 + \frac{4 \pi e^2}{k^2 +
k^2_{TF}}>0,
\end{equation}
where
\begin{eqnarray}
V_0 &=& \frac{\partial \mu_p}{\partial \rho_p} - \frac{(\partial
\mu_n / \partial \rho_p)^2}{\partial \mu_n / \partial \rho_n},
\label{v0}\\
\beta &=& D_{pp} + 2 D_{np} \zeta + D_{nn} \zeta^2,~~\zeta =
-\frac{\partial \mu_p / \partial \rho_n}{\partial \mu_n / \partial \rho_n},\\
k^2_{TF} &=& \frac{4 \pi e^2}{\partial \mu_e / \rho_e}.
\end{eqnarray}
In the above expressions, we used the relation $\frac{\partial \mu
_{n}}{\partial \rho _{p}}=\frac{\partial \mu _{p}}{\partial \rho
_{n}}$ following $\frac{\partial \mu _{n}}{\partial \rho _{p}}=\frac{\partial }{%
\partial \rho _{p}}\left( \frac{\partial \varepsilon }{\partial \rho _{n}}%
\right) =\frac{\partial }{\partial \rho _{n}}\left( \frac{\partial
\varepsilon }{\partial \rho _{p}}\right) =\frac{\partial \mu
_{p}}{\partial \rho _{n}}$ with $\varepsilon $ being the energy
density of the $npe$ matter. Meanwhile, $\frac{\partial
\mu_n}{\partial \rho_n}$ is assumed to be positive. If we have
$\frac{\partial \mu_n}{\partial \rho_n}<0$ but $\frac{\partial
\mu_p}{\partial \rho_p}>0$ we can change the form of the equations
correspondingly. In this form, it is clear that the density gradient
and the Coulomb term clearly contribute positively to the
$V_{dyn}(k)$. They thus help to make the system more stable. At $k =
[ (\frac{4 \pi e^2}{\beta})^{1/2} - k^2_{TF} ]^{1/2}$, the
$V_{dyn}(k)$ has a minimal value of~\citep{BBP71,Pet95b}
\begin{equation}\label{Vdynmin}
V_{dyn} = V_0 + 2 (4 \pi e^2 \beta)^{1/2} - \beta k^2_{TF}.
\end{equation}
Then the density at which Eq.~(\ref{Vdynmin}) becomes zero
determines the instability boundary.

\subsection{The thermodynamical method}
\label{thermodynamical}

The thermodynamical method requires the system to obey the stability
condition~\citep{Kub07,Lat07}
\begin{equation}\label{ther1}
-\left(\frac{\partial P}{\partial v}\right)_\mu>0,
\end{equation}
\begin{equation}\label{ther2}
-\left(\frac{\partial \mu}{\partial q_c}\right)_v>0,
\end{equation}
or the system will be unstable against small density fluctuations.
These conditions are equivalent to requiring the convexity of the
energy per particle in the single phase~\citep{Kub07,Lat07} by
ignoring the finite size effects due to surface and Coulomb energies
as shown in the following. Here the $P=P_b+P_e$ is the total
pressure of the $npe$ system with the contributions $P_b$ and $P_e$
from baryons and electrons, respectively. The $v$ and $q_c$ are the
volume and charge per baryon number. The $\mu$ is the chemical
potential defined as
\begin{equation}
\mu=\mu_n-\mu_p.
\end{equation}
In fact, Eq.~(\ref{ther1}) is simply the well-known mechanical
stability condition of the system at a fixed $\mu$. It ensures that
any local density fluctuation will not diverge. On the other hand,
Eq.~(\ref{ther2}) is the charge or chemical stability condition of
the system at a fixed density. It means that any local charge
variation violating the charge neutrality condition will not
diverge. If the $\beta$-equilibrium condition is satisfied, namely
$\mu=\mu_e$, the electron contribution to the pressure $P_e$ is only
a function of the chemical potential $\mu$, and in this case one can
rewrite Eq.~(\ref{ther1}) as
\begin{equation}
-\left(\frac{\partial P_b}{\partial v}\right)_\mu>0.
\end{equation}
By using the relation $\frac{\partial E_b(\rho,x_p)}{\partial x_p}
= -\mu$, one can get~\citep{Kub07}
\begin{eqnarray}\label{ther3}
- \left(\frac{\partial P_b}{\partial v}\right)_\mu =
2\rho^3\frac{\partial E_b(\rho,x_p)}{\partial \rho} + \rho^4
\frac{\partial^2 E_b(\rho,x_p)}{\partial \rho^2}- \rho^4
\left(\frac{\partial^2 E_b(\rho,x_p)}{\partial \rho
\partial x_p}\right)^2/\frac{\partial^2 E_b(\rho,x_p)}{\partial x_p^2},
\end{eqnarray}
\begin{equation}\label{ther4}
-\left(\frac{\partial q_c}{\partial \mu}\right)_v =
1/\frac{\partial^2 E_b(\rho,x_p)}{\partial x_p^2} + \frac{\partial
\rho_e}{\partial \mu_e}/\rho,
\end{equation}
where $q_c=x_p-\rho_e/\rho$. The $\rho=1/v$ is the baryon density
and the $E_b(\rho,x_p)$ is the energy per baryon. Within the free
Fermi gas model, the density of electrons $\rho_e$ is uniquely
determined by the electron chemical potential $\mu_e$. Then the
thermodynamical relations Eq.~(\ref{ther1}) and Eq.~(\ref{ther2})
are identical to~\citep{Lat07,Kub07}
\begin{eqnarray}\label{ther5}
-\left(\frac{\partial P_b}{\partial v}\right)_\mu = \rho^2 \left[2
\rho \frac{\partial E_b(\rho,x_p)}{\partial \rho} + \rho^2
\frac{\partial^2 E_b(\rho,x_p)}{\partial \rho^2} -
\left(\frac{\partial^2 E_b(\rho,x_p)}{\partial \rho
\partial x_p}\rho\right)^2/\frac{\partial^2 E_b(\rho,x_p)}{\partial x_p^2}\right]>0,
\end{eqnarray}
\begin{equation}\label{ther6}
-\left(\frac{\partial q_c}{\partial \mu}\right)_v =
1/\frac{\partial^2 E_b(\rho,x_p)}{\partial
x_p^2}+\frac{\mu^2_e}{\pi^2\hbar^3\rho}>0,
\end{equation}
respectively. The second inequality is usually valid. Thus, the following condition from the first
one
\begin{eqnarray} \label{Vther}
V_{ther} = 2 \rho \frac{\partial E_b(\rho,x_p)}{\partial \rho} +
\rho^2 \frac{\partial^2 E_b(\rho,x_p)}{\partial \rho^2} -
\left(\frac{\partial^2 E_b(\rho,x_p)}{\partial \rho
\partial x_p}\rho\right)^2/\frac{\partial^2 E_b(\rho,x_p)}{\partial x_p^2}
\end{eqnarray}
determines the thermodynamical instability region.

Within the parabolic approximation neglecting higher order terms of
isospin asymmetry $\delta=1-2x_p$, the EOS of asymmetric nuclear
matter is
\begin{equation}
E_b(\rho,\delta)=E_0(\rho)+E_{sym}(\rho)\delta^2,
\end{equation}
where $E_0(\rho)$ is the energy per nucleon of symmetric nuclear
matter. Then Eq.~(\ref{Vther}) can be reexpressed as~\citep{Lat07}
\begin{eqnarray}\label{Vtherpa}
V^{PA}_{ther} = \rho^2 \frac{d^2 E_0}{d \rho^2} + 2 \rho \frac{d
E_0}{d \rho} + (1-2x_p)^2 \left[ \rho^2 \frac{d^2 E_{sym}}{d \rho^2}
+ 2 \rho \frac{d E_{sym}}{d \rho} - 2 E^{-1}_{sym}
\left(\rho \frac{d E_{sym}}{d \rho}\right)^2\right]. \notag\\
\end{eqnarray}

\subsection{The relationship between the dynamical and thermodynamical methods}
\label{relation}

The Eq.~(\ref{Vdynmin}) and Eq.~(\ref{Vther}) together with the
relationship between the density $\rho$ and the proton fraction
$x_p$ required by the $\beta$-equilibrium and the charge neutrality
conditions will then determine respectively the dynamical and the
thermodynamical core-crust transition density in neutron stars.
These two methods together with various EOS's have been widely used
in the literature while their relationship is still unclear.
Therefore, it would be interesting to first obtain some analytical
insights on their relationship before comparing their numerical
predictions.

In the following, we first analyze the instability of asymmetric
nuclear matter without considering the $\beta$-equilibrium and the
charge neutrality conditions. As used in the previous subsection
and the literature, the stability condition is often expressed
using the $\rho$ and $x_p$ within the dynamical method while the
$\rho_n=(1-x_p)\rho$ and $\rho_p=x_p \rho$ within the
thermodynamical one, respectively. Thus the following simple
thermodynamical relations are useful for understanding the
relationship between the two methods
\begin{equation}\label{best}
\frac{\partial E_b}{\partial x_p} = -\mu = \mu_p - \mu_n,
\end{equation}
\begin{equation}
\frac{\partial P_b}{\partial \rho}=(1-x_p) \rho \frac{\partial
\mu_n}{\partial \rho}+x_p \rho \frac{\partial \mu_p}{\partial \rho},
\end{equation}
\begin{equation}
\frac{\partial P_b}{\partial x_p}=(1-x_p) \rho \frac{\partial
\mu_n}{\partial x_p} + x_p \rho \frac{\partial \mu_p}{\partial x_p},
\end{equation}
where the pressure of baryons is
$P_b=\mu_n\rho_n+\mu_p\rho_p-E_b\rho$. In this way, the derivatives
of the energy of baryons can be expressed as
\begin{eqnarray}
\frac{\partial E_b}{\partial \rho} &=& \frac{P_b}{\rho^2}, \label{dEdrho}\\
\frac{\partial^2 E_b}{\partial \rho^2} &=& \frac{\partial}{\partial
\rho}\left(\frac{P_b}{\rho^2}\right) \notag\\
&=& -\frac{2 P_b}{\rho^3}+\frac{1}{\rho^2}\left[(1-x_p) \rho
\frac{\partial \mu_n}{\partial \rho} + x_p \rho \frac{\partial
\mu_p}{\partial
\rho}\right] \notag\\
&=& - \frac{2 P_b}{\rho^3} + \frac{1}{\rho^2} \left[(1-x_p)^2 \rho
\frac{\partial \mu_n}{\partial \rho_n} +
x_p (1-x_p) \rho \frac{\partial \mu_n}{\partial \rho_p} \right] \notag \\
&+& \frac{1}{\rho^2}\left[ x_p (1-x_p) \rho \frac{\partial
\mu_p}{\partial \rho_n} + x_p^2 \rho\frac{\partial \mu_p}{\partial
\rho_p} \right], \label{d2Edrho2}\\
\frac{\partial^2 E_b}{\partial x_p^2} &=& - \frac{\partial
\mu}{\partial x_p}=\frac{\partial \mu_p}{\partial x_p} -
\frac{\partial \mu_n}{\partial
x_p} \notag\\
&=& \rho \left(\frac{\partial \mu_p}{\partial \rho_p}-\frac{\partial
\mu_p}{\partial \rho_n}-\frac{\partial \mu_n}{\partial
\rho_p}+\frac{\partial \mu_n}{\partial \rho_n}\right), \label{d2EdX2}\\
\frac{\partial^2 E_b}{\partial \rho \partial x_p} &=&
-\frac{\partial \mu}{\partial \rho} = \frac{\partial \mu_p}{\partial
\rho} - \frac{\partial \mu_n}{\partial
\rho} \notag\\
&=& (1-x_p) \frac{\partial \mu_p}{\partial \rho_n} +
x_p\frac{\partial \mu_p}{\partial \rho_p} - (1-x_p) \frac{\partial
\mu_n}{\partial \rho_n} - x_p\frac{\partial \mu_n}{\partial \rho_p}.
\label{d2EdrhoX}\\\notag
\end{eqnarray}
As shown earlier, for nuclear matter without considering the
Coulomb interaction one has
\begin{equation}\label{npsymm}
\frac{\partial \mu_n}{\partial \rho_p}=\frac{\partial
\mu_p}{\partial \rho_n}.
\end{equation}
From Eq.~(\ref{dEdrho}), (\ref{d2Edrho2}), (\ref{d2EdX2}),
(\ref{d2EdrhoX}) and (\ref{npsymm}), we can then obtain the
following important equality
\begin{equation}\label{relation2}
\frac{2}{\rho}\frac{\partial E_b}{\partial \rho}\frac{\partial^2
E_b}{\partial x_p^2}+\frac{\partial^2 E_b}{\partial
\rho^2}\frac{\partial^2 E_b}{\partial
x_p^2}-\left(\frac{\partial^2 E_b}{\partial \rho \partial
x_p}\right)^2= \frac{\partial \mu_n}{\partial
\rho_n}\frac{\partial \mu_p}{\partial \rho_p}-\left(\frac{\partial
\mu_n}{\partial \rho_p}\right)^2.
\end{equation}
Therefore, for positive values of $\frac{\partial^2 E_b}{\partial
x_p^2}$, the condition Eq.~(\ref{ther5}) is simply equivalent to
requiring a positive bulk term $V_{0}$ in the Eq.~(\ref{dymethod}).
Since the transition density is usually in the sub-saturation
density region where the $\frac{\partial^2 E_b}{\partial x_p^2}>0$
is valid for almost all model EOS's, the thermodynamical stability
condition is thus simply the limit of the dynamical one as
$k\rightarrow 0$ (long-wavelength limit) when the Coulomb
interaction is neglected.

\section{The EOS and symmetry energy with selected 51 Skyrme forces and a modified Gogny interaction}
\label{EOS}

In this section, we summarize the EOS and the corresponding symmetry
energy obtained using the modified finite-range Gogny effective
interaction (MDI)~\citep{Das03} and $51$ popular Skyrme forces
within the Hartree-Fock approach. These results will be used later
in our numerical calculations of the core-crust transition density
and pressure. The MDI interaction has been extensively used in our
previous studies of heavy-ion collisions, the liquid-gas phase
transition in asymmetric nuclear matter and several issues in
astrophysics~\citep{LCK08}. The EOS's using various Skyrme forces
are well known for their simple forms and successful descriptions of
many interesting phenomenon, see, e.g.,
refs.~\citep{Chabanat,Ste05,Stone,Sto07}. A very useful feature of
both the MDI and the Skyrme interaction is that analytical
expressions for many interesting physical quantities in asymmetric
nuclear matter at zero temperature can be obtained.

\begin{table}[tbp]
\caption{{\protect\small Saturation density $\rho_0$ (fm$^{-3}$),
binding energy of symmetric nuclear matter $E_0(\rho_0)$ (MeV),
incompressibility $K_0$ (MeV), symmetry energy $E_{sym}(\rho_0)$
(MeV) as well as slope and curvature parameters of symmetry energy
$L$ (MeV) and $K_{sym}$ (MeV) at the saturation density.}}
\label{SHFtab1}
\begin{tabular}{ccccccc}
\hline\hline
SHF & $\rho_0$ & \quad $E_0(\rho_0)$ \quad & $K_0$ & \quad $E_{sym}(\rho_0)$ & \quad $L$ & \quad $K_{sym}$ \quad \\
\hline
$$ BSk3  &  0.157 & -15.8 &   234.8 & 27.9 & 6.8 & -306.9 \\
$$ BSk1  &  0.157 & -15.8 &   231.3 & 27.8 & 7.2 & -281.8 \\
$$ BSk2  &  0.157 & -15.8 &   233.7 & 28.0 & 8.0 & -297.0 \\
$$ MSk7  &  0.157 & -15.8 &   231.2 & 27.9 & 9.4 & -274.6 \\
$$ BSk4  &  0.157 & -15.8 &   236.8 & 28.0 & 12.5 & -265.9 \\
$$ BSk8  &  0.159 & -15.8 &   230.3 & 28.0 & 14.9 & -220.9 \\
$$ BSk6  &  0.157 & -15.8 &   229.1 & 28.0 & 16.8 & -215.2 \\
$$ BSk7  &  0.157 & -15.8 &   229.3 & 28.0 & 18.0 & -209.4 \\
$$ SKP & 0.163 & -16.0  &  201.0 & 30.0 & 19.6 & -266.8 \\
$$ BSk5  &  0.157 & -15.8 &   237.2 & 28.7 & 21.4 & -240.3 \\
$$ SKXm  &  0.159 & -16.0  &  238.1 & 31.2 & 32.1 & -242.8 \\
$$ RATP  &  0.160  & -16.0  &  239.4 & 29.2 & 32.4 & -191.2 \\
$$ SKX  & 0.155 & -16.1  &  271.1 & 31.1 & 33.2 & -252.1 \\
$$ SKXce  & 0.155 & -15.9  &  268.2 & 30.1 & 33.5 & -238.4 \\
$$ BSk15  &  0.159 & -16.0 &   241.6 & 30.0 & 33.6 & -194.3 \\
$$ BSk16  &  0.159 & -16.1 &   241.7 & 30.0 & 34.9 & -187.4 \\
$$ BSk10  &  0.159 & -15.9 &   238.8 & 30.0 & 37.2 & -194.9 \\
$$ SGII  &  0.158 & -15.6  &  214.7 & 26.8 & 37.6 & -145.9 \\
$$ BSk12  &  0.159 & -15.9 &   238.1 & 30.0 & 38.0 & -191.4 \\
$$ BSk11  &  0.159 & -15.9 &   238.1 & 30.0 & 38.4 & -189.8 \\
$$ BSk13  &  0.159 & -15.9 &   238.1 & 30.0 & 38.8 & -187.9 \\
$$ BSk9  &  0.159 & -15.9 &   231.4 & 30.0 & 39.9 & -145.3 \\
$$ SLy10 &  0.158 & -16.5  &  237.8 & 33.2 & 40.8 & -148.0 \\
$$ BSk14  &  0.159 & -15.9 &   239.3 & 30.0 & 43.9 & -152.0 \\
$$ SLy230a & 0.160 & -16.0  &  229.9 & 32.0 & 44.3 & -98.2 \\
$$ SKM$^\star$  &  0.160 & -15.8  &  216.6 & 30.0 & 45.8 & -155.9 \\

\hline\hline
\end{tabular}%
\end{table}

\begin{table}[tbp]
\caption{{\protect\small Continued with Table \ref{SHFtab1}}}
\label{SHFtab2}
\begin{tabular}{ccccccc}
\hline\hline
SHF & $\rho_0$ & \quad $E_0(\rho_0)$\quad & $K_0$ & \quad $E_{sym}(\rho_0)$ & \quad $L$ & \quad $K_{sym}$ \quad\\
\hline
$$ SLy230b & 0.160 & -16.0  &  229.9 & 32.0 & 46.0 & -119.7 \\
$$ SLy6  &  0.161 & -16.5  &  237.9 & 32.2 & 46.7 & -117.0 \\
$$ SLy8  &  0.163 & -16.6  &  238.0 & 32.4 & 46.8 & -121.0 \\
$$ SLy4  &  0.162 & -16.6  &  238.0 & 32.8 & 46.9 & -124.6 \\
$$ SLy0  &  0.163 & -16.6  &  238.3 & 32.4 & 46.9 & -121.4 \\
$$ SLy3  &  0.163 & -16.6  &  238.0 & 33.1 & 47.0 & -126.9 \\
$$ SKM & 0.160 & -15.8 &   216.6 & 30.7 & 49.3 & -148.8 \\
$$ SLy7  &  0.161 & -16.5 & 237.8 & 33.4 & 49.7 & -118.9 \\
$$ SLy2  &  0.162 & -16.5 &   237.3 & 33.3 & 50.3 & -117.9 \\
$$ SLy1  &  0.163 & -16.6  &  237.9 & 33.5 & 50.4 & -120.2 \\
$$ SLy5  &  0.163 & -16.6  &  238.0 & 33.6 & 51.9 & -116.3 \\
$$ SLy9  &  0.153 & -16.4  &   237.7 & 33.2 & 57.2 & -84.9 \\
$$ SkI6  &  0.159 & -15.9 &   248.2 & 29.9 & 59.2 & -46.8 \\
$$ SkI4  &  0.160 & -15.9  &  248.0 & 29.5 & 60.4 & -40.6 \\
$$ SGI & 0.154 & -15.9 &   261.8 & 28.3 & 63.9 & -52.0 \\
$$ SKO$^\star$  &  0.160 & -15.7  &  222.1 & 32.1 & 69.7 & -77.5 \\
$$ SkMP  &  0.159 & -16.1  &  238.5 & 30.1 & 70.7 & -51.4 \\
$$ SKa & 0.155 & -16.0 &   263.2 & 32.9 & 74.6 & -78.5 \\
$$ SKO & 0.160 & -15.8 &   222.8 & 32.0 & 79.5 & -42.3 \\
$$ R$_{\sigma}$ & 0.158 & -15.6  &  237.4 & 30.6 & 85.7 & -9.1 \\
$$ SKT4  &  0.157 & -15.5  &  229.3 & 34.8 & 92.4 & -24.2 \\
$$ G$_{\sigma}$ & 0.158 & -15.6 &   237.2 & 31.4 & 94.0 & 14.0 \\
$$ SkI3  &  0.158 & -16.0 &   258.2 & 34.8 & 100.5 & 73.0 \\
$$ SkI2  &  0.158 & -15.8 &   240.9 & 33.4 & 104.3 & 70.7 \\
$$ SkI5  &  0.156 & -15.8 &   255.8 & 36.6 & 129.3 & 159.6 \\
\hline\hline
\end{tabular}%
\end{table}

\subsection{The EOS and symmetry energy with selected 51 Skyrme interactions}
\label{SHF}

Within the SHF approach the energy per nucleon for symmetric
nuclear matter can be expressed as~\citep{Chabanat}
\begin{eqnarray}\label{SHFE0}
E_0(\rho) = \frac{3 \hbar^2}{10 m}
\left(\frac{3\pi^2}{2}\right)^{2/3} \rho^{2/3} + \frac{3}{8} t_0
\rho + \frac{3}{80} \Theta_s \left(\frac{3\pi^2}{2}\right)^{2/3}
\rho^{5/3} + \frac{1}{16} t_3 \rho^{\sigma + 1},
\end{eqnarray}
with $\Theta_s=3t_1+(5+4x_2)t_2$. For asymmetric nuclear matter, the
energy per nucleon is~\citep{Chabanat}
\begin{eqnarray}\label{SHFE}
E_b(\rho,\delta~or~x_p) &=& \frac{3 \hbar^2}{10 m}
\left(\frac{3\pi^2}{2}\right)^{2/3} \rho^{2/3} F_{5/3} + \frac{1}{8}
t_0 \rho [2 (x_0 + 2) - (2 x_0 + 1) F_2] \notag\\
&+& \frac{1}{48} t_3 \rho^{\sigma + 1} [2 (x_3 + 2) - (2 x_3 + 1)
F_2] + \frac{3}{40} \left(\frac{3\pi^2}{2}\right)^{2/3} \rho^{5/3}
\notag \\
&\times& \left\{ [t_1 (x_1 + 2) + t_2 (x_2 + 2)] F_{5/3} +
\frac{1}{2} [t_2 (2 x_2 + 1) - t_1 (2 x_1 + 1)]
F_{8/3} \right\},\notag\\
\end{eqnarray}
with
\begin{equation}
F_m(\delta) = \frac{1}{2} [(1 + \delta)^m + (1 - \delta)^m], \notag
\end{equation}
\begin{equation}
F_m(x_p) = 2^{m - 1} [x_p^m + (1 - x_p)^m] \notag.
\end{equation}
Within the parabolic approximation widely used in the literature,
the symmetry energy is calculated from
\begin{equation}\label{Esym-pa}
E_{sym}(\rho)\approx E_b(\rho,\delta =1)-E_b(\rho,\delta =0).
\end{equation}
But strictly speaking, the symmetry energy should be the
coefficient of $\delta^2$ in the Taylor expansion of
$E_b(\rho,\delta)$ in terms of $\delta$, i.e.,
\begin{eqnarray}\label{Esym-d}
E_{sym}(\rho) &=& \frac{1}{2} \left(\frac{\partial^2 E_b}{\partial
\delta^2}\right)_{\delta=0}.
\end{eqnarray}
We notice here that the above two definitions for the symmetry
energy would be the same should there be no higher order terms in
$\delta$ in the EOS of asymmetric nuclear matter (But it should be
noted that the kinetic part of the EOS of asymmetric nuclear matter
always contains higher order terms in $\delta$).

Thus, by definition of Eq.~(\ref{Esym-d}), for Skyrme interactions,
one has
\begin{eqnarray}\label{SHFEsym}
E_{sym}(\rho) &=& \frac{1}{2} \left(\frac{\partial^2 E_b}{\partial
\delta^2}\right)_{\delta=0} \notag\\
&=& \frac{\hbar^2}{6 m} \left(\frac{3\pi^2}{2}\right)^{2/3}
\rho^{2/3} - \frac{1}{8} t_0 (2 x_0 + 1) \rho \notag\\
&-& \frac{1}{24} \left(\frac{3\pi^2}{2}\right)^{2/3} \Theta_{sym}
\rho^{5/3} - \frac{1}{48} t_3 (2 x_3 + 1) \rho^{\sigma + 1},
\end{eqnarray}
where $\Theta_{sym}=3t_1x_1-t_2(4+5x_2)$. $\sigma$, $t_0 \sim t_3$
and $x_0 \sim x_3$ are the Skyrme parameters.

As it has been used extensively by many authors, near the saturation
density $\rho_0$ the symmetry energy can be expanded as
\begin{equation}
E_{sym}(\rho) \approx E_{sym}(\rho_0) + \frac{L}{3}
\left(\frac{\rho-\rho_0}{\rho_0}\right) + \frac{K_{sym}}{18}
\left(\frac{\rho-\rho_0}{\rho_0}\right)^2,
\end{equation}
where $L$ and $K_{sym}$ are, respectively, the slope parameter and
curvature parameter of the symmetry energy at $\rho_0$, i.e.,
\begin{eqnarray}
L &=& 3\rho_0 \frac{\partial E_{sym}(\rho)}{\partial
\rho}|_{\rho=\rho_0},\\ \label{L} K_{sym} &=&
9\rho^2_0\frac{\partial^2 E_{sym}(\rho)}{\partial
\rho^2}|_{\rho=\rho_0}. \label{K}
\end{eqnarray}
The $L$ and $K_{sym}$ can be used conveniently to characterize the
density dependence of the symmetry energy around the saturation
density $\rho_0$. In the present work we use $51$ standard Skyrme
forces with their saturation density and the symmetry energy
satisfying $0.140$ fm$^{-3}<\rho_0<0.165$ fm$^{-3}$ and $26$
MeV$<E_{sym}(\rho_0)<37$ MeV, respectively. Some Skyrme forces with
very small or negative $L$ values are not considered here as they
generally predict bound pure neutron matter at sub-saturation
densities and are not suitable for the description of neutron-rich
environments like neutron star crusts as discussed in detail by
Stone et al.~\citep{Stone}. In addition, we have not included the
Skyrme forces predicting values for the incompressibility $K_0$
inconsistent with the empirical value of about $240 \pm 40$ MeV. The
detailed values of the parameters for these $51$ Skyrme forces can
be found in
refs.~\citep{Bra85,Fri86,Bro98,Chabanat,Stone,Sto07,Che05b,Ste05b,%
Sam02,Sam03,Gor03,Sam04,Gor05,Sam05,Gor06,Gor07,Cha08a,Gor08}. The
selected ranges of $\rho_0$ and $E_{sym}(\rho_0)$ are consistent
with their empirical values inferred from nuclear laboratory data.
The detailed properties of these forces at $\rho_0$ are summarized
in Tables~\ref{SHFtab1} and ~\ref{SHFtab2} in the order of rising
values of $L$.

\subsection{The EOS and symmetry energy with the modified Gogny interaction MDI}
\label{MDI}

For the MDI interaction based on the Hartree-Fock calculation using
the Gogny interaction, the baryon potential energy density can be
expressed as~\citep{Das03}
\begin{eqnarray}
V(\rho,\delta ) &=&\frac{A_{u}(x)\rho _{n}\rho _{p}}{\rho _{0}}
+\frac{A_{l}(x)}{2\rho _{0}}(\rho _{n}^{2}+\rho
_{p}^{2})+\frac{B}{\sigma +1}\frac{\rho ^{\sigma +1}}{\rho
_{0}^{\sigma }} (1-x\delta ^{2})
\notag\\
&+&\frac{1}{\rho _{0}}\sum_{\tau ,\tau ^{\prime
}}C_{\tau ,\tau ^{\prime }} \int \int d^{3}pd^{3}p^{\prime }\frac{f_{\tau }(\vec{r},\vec{p}%
)f_{\tau ^{\prime }}(\vec{r},\vec{p}^{\prime
})}{1+(\vec{p}-\vec{p}^{\prime })^{2}/\Lambda ^{2}}.  \label{MDIVB}
\end{eqnarray}%
We notice here that the above is a natural extension to isospin
asymmetric case of the corresponding potential energy density for
symmetric nuclear matter given in
refs.\citep{GBD87,Pra88,Wel88,Gal90}. It is similar to the BGBD
(Bombaci-Gale-Bertsch-Das Gupta) potential energy
density~\citep{Bom01}. The MDI interaction has been used extensively
in studying heavy-ion reactions~\citep{LCK08}, liquid-gas phase
transitions in neutron-rich matter~\citep{Xu07a,LiBA07,Xu07b} and
several structural properties~\citep{Lis06,Kra07,Kra08a} and
gravitational wave emissions~\citep{Kra08b,Wor09} of neutron stars.

In the mean field approximation, Eq.~(\ref{MDIVB}) leads to the
following single particle potential for a nucleon with momentum $\vec{p}$ and isospin $%
\tau $, i.e.,
\begin{eqnarray}
U(\rho,\delta ,\vec{p},\tau ) &=&A_{u}(x)\frac{\rho _{-\tau }}{\rho _{0}}%
+A_{l}(x)\frac{\rho _{\tau }}{\rho _{0}}
+B(\frac{\rho }{\rho _{0}})^{\sigma }(1-x\delta ^{2})-8\tau x\frac{B}{%
\sigma +1}\frac{\rho ^{\sigma -1}}{\rho _{0}^{\sigma }}\delta \rho
_{-\tau }
\notag \\
&+&\frac{2C_{\tau ,\tau }}{\rho _{0}}\int d^{3}p^{\prime }\frac{f_{\tau }(%
\vec{r},\vec{p}^{\prime })}{1+(\vec{p}-\vec{p}^{\prime
})^{2}/\Lambda ^{2}}
+\frac{2C_{\tau ,-\tau }}{\rho _{0}}\int d^{3}p^{\prime }\frac{f_{-\tau }(%
\vec{r},\vec{p}^{\prime })}{1+(\vec{p}-\vec{p}^{\prime
})^{2}/\Lambda ^{2}}. \label{MDIU}
\end{eqnarray}%
In the above the isospin $\tau =1/2$ ($-1/2$) for neutrons
(protons). The coefficients $A_{u}(x)$ and $A_{l}(x)$
are~\citep{Che05a}
\begin{equation}\label{alau}
A_{u}(x)=-95.98-x\frac{2B}{\sigma
+1},~~~~A_{l}(x)=-120.57+x\frac{2B}{\sigma +1}.
\end{equation}
The values of the parameters are $\sigma=4/3$, $B=106.35$ MeV,
$C_{\tau ,\tau }=-11.70$ MeV, $C_{\tau ,-\tau }=-103.40$ MeV and
$\Lambda=p_f^{0}$ which is the Fermi momentum of nuclear matter at
$\rho_0$~\citep{Das03}. The parameter $x$ was introduced to mimic
various $E_{sym}(\rho)$ predicted by different microscopic many-body
theories. By adjusting the $x$ parameter, the $E_{sym}(\rho)$ is
varied without changing any property of symmetric nuclear matter and
the symmetry energy at saturation density as the $x$-dependent
$A_{u}(x)$ and $A_{l}(x)$ are automatically adjusted accordingly. We
note especially that the symmetry energy at normal density
$E_{sym}(\rho_0)$ is fixed independent of the $x$ parameter. Using
the definition in Eq.~ (\ref{Esym-d}), $E_{sym}(\rho_0)=30.54$ MeV
at $\rho _{0}=0.16$ fm$^{-3}$ while its value is 31.6 MeV within the
parabolic approximation of Eq.~(\ref{Esym-pa}).

\begin{figure}[tbh]
\centering
\includegraphics[scale=1.2]{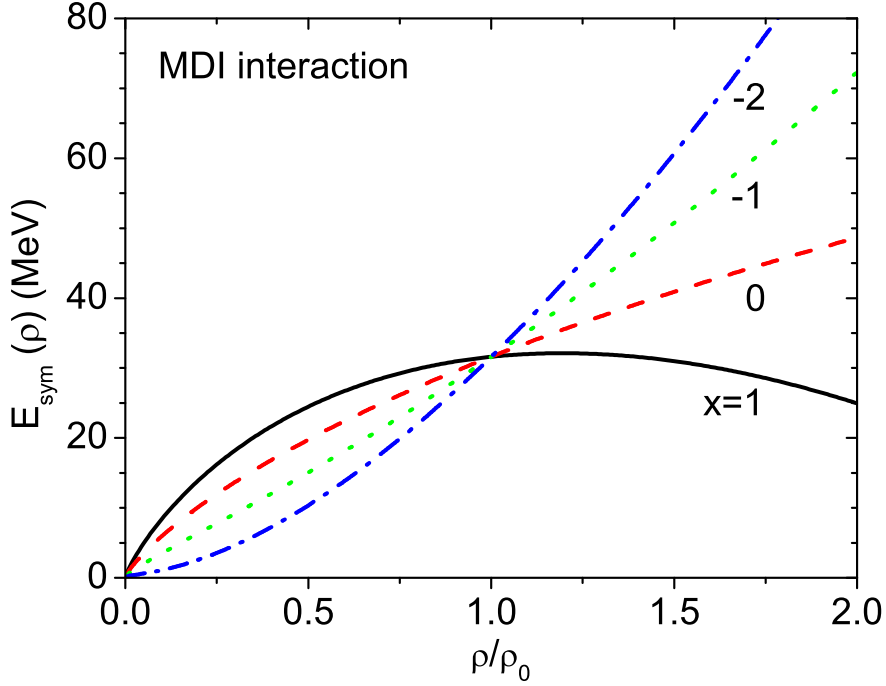}
\caption{(Color online) The density dependence of the nuclear
symmetry energy for different values of the parameter $x$ in the MDI
interaction. Taken from ref. \citep{Che05a}} \label{MDIsymE}
\end{figure}

At zero temperature the phase space distribution function can be
written as $f_{\tau }(\vec{r},\vec{p})$ $=\frac{2}{h^{3}}\Theta
(p_{f}(\tau )-p)$, and all the integrals expressions can be
calculated analytically~\citep{Wel88,Das03,Chen07}
\begin{eqnarray}
&\int& d^{3}p^{\prime }\frac{f_{\tau }(%
\vec{r},\vec{p}^{\prime })}{1+(\vec{p}-\vec{p}^{\prime
})^{2}/\Lambda ^{2}} \notag \\
&=& \frac{2}{h^3} \pi \Lambda^3 [\frac{p^2_f(\tau) + \Lambda^2 -
p^2}{2 p \Lambda} \ln \frac{[p + p_f(\tau)]^2 + \Lambda^2}{[p -
p_f(\tau)]^2 + \Lambda^2} \notag\\
&+& \frac{2 p_f(\tau)}{\Lambda} - 2 \{\arctan\frac{p +
p_f(\tau)}{\Lambda} - \arctan\frac{p - p_f(\tau)}{\Lambda}\}],\notag\\
\end{eqnarray}
\begin{eqnarray}
&\int \int& d^{3}pd^{3}p^{\prime }\frac{f_{\tau }(\vec{r},\vec{p}%
)f_{\tau ^{\prime }}(\vec{r},\vec{p}^{\prime
})}{1+(\vec{p}-\vec{p}^{\prime })^{2}/\Lambda ^{2}} \notag\\
&=& \frac{1}{6} \left(\frac{4 \pi}{h^3}\right)^2 \Lambda^2 \{
p_f(\tau) p_f(\tau^\prime) [3 (p^2_f(\tau) + p^2_f(\tau^\prime)) -
\Lambda^2] \notag\\
&+& 4 \Lambda [(p^3_f(\tau) - p^3_f(\tau^\prime))
\arctan\frac{p_f(\tau) - p_f(\tau^\prime)}{\Lambda} \notag \\
&-& (p^3_f(\tau) + p^3_f(\tau^\prime)) \arctan\frac{p_f(\tau) +
p_f(\tau^\prime)}{\Lambda}] \notag \\
&+& \frac{1}{4}[\Lambda^4 + 6 \Lambda^2 (p^2_f(\tau) +
p^2_f(\tau^\prime)) - 3 (p^2_f(\tau) - p^2_f(\tau^\prime))^2]
\notag\\
&\times& \ln \frac{(p_f(\tau) + p_f(\tau^\prime))^2 +
\Lambda^2}{(p_f(\tau) - p_f(\tau^\prime))^2 + \Lambda^2} \}.
\end{eqnarray}
The kinetic energy is
\begin{eqnarray}
E_k(\rho,\delta) &=& \frac{1}{\rho}\int d^{3}p \left(\frac{p^{2}}{2
m}
f_n(\vec{r},\vec{p}) + \frac{p^{2}}{2m} f_p(\vec{r},\vec{p})\right) \notag \\
&=& \frac{4 \pi}{5 m h^3 \rho} (p^5_{fn} + p^5_{fp}),
\end{eqnarray}
where $p_{fn(p)}=\hbar (3 \pi^2 \rho_{n(p)})^{1/3}$ is the Fermi
momentum of neutrons(protons). Then, the total energy per baryon for cold asymmetric nuclear
matter is
\begin{equation}
E_b(\rho,\delta) = \frac{V(\rho,\delta )}{\rho} +
E_k(\rho,\delta).
\end{equation}
By setting $\rho_n=\rho_p=\frac{\rho}{2}$ and $p_{fn}=p_{fp}=p_f$ we
thus obtain the following EOS of cold symmetric nuclear
matter
\begin{eqnarray}
E_0(\rho) &=& \frac{8 \pi}{5 m h^3 \rho} p^5_f
+ \frac{\rho}{4 \rho_0} (A_l(x) + A_u(x)) \notag\\
&+& \frac{B}{\sigma + 1} \left(\frac{\rho}{\rho_0}\right)^\sigma +
\frac{1}{3 \rho_0 \rho}
(C_l + C_u) \left(\frac{4 \pi}{h^3}\right)^2 \Lambda^2 \notag\\
&\times& \left[p^2_f (6 p^2_f - \Lambda^2) - 8 \Lambda p^3_f \arctan
\frac{2 p_f}{\Lambda} + \frac{1}{4} (\Lambda^4 + 12 \Lambda^2 p^2_f)
\ln \frac{4 p^2_f + \Lambda^2}{\Lambda^2}\right].
\end{eqnarray}
We stress here that since the $A_l(x) + A_u(x)$ is a constant of
$-216.55$ MeV according to Eq.~(\ref{alau}), the $E_0(\rho)$ is
independent of the parameter $x$ as expected. The symmetry energy by
definition is
\begin{eqnarray}\label{esymmdi}
E_{sym}(\rho) &=& \frac{1}{2} \left(\frac{\partial^2 E}
{\partial \delta^2}\right)_{\delta=0} \notag\\
&=& \frac{8 \pi}{9 m h^3 \rho} p^5_f + \frac{\rho}{4 \rho_0} (A_l(x)-
A_u(x)) - \frac{B x}{\sigma + 1}
\left(\frac{\rho}{\rho_0}\right)^\sigma \notag\\
&+& \frac{C_l}{9 \rho_0 \rho} \left(\frac{4 \pi}{h^3}\right)^2
\Lambda^2 \left[4 p^4_f - \Lambda^2 p^2_f \ln \frac{4 p^2_f
+ \Lambda^2}{\Lambda^2}\right] \notag\\
&+& \frac{C_u}{9 \rho_0 \rho} \left(\frac{4 \pi}{h^3}\right)^2
\Lambda^2 \left[4 p^4_f - p^2_f (4 p^2_f + \Lambda^2) \ln \frac{4
p^2_f + \Lambda^2}{\Lambda^2}\right],
\end{eqnarray}
where $p_f=\hbar(3\pi^2\frac{\rho}{2})^{1/3}$ is the Fermi momentum
for symmetric nuclear matter. We note here that since the
$A_l(x)-A_u(x)=-24.59+4Bx/(\sigma +1)$ according to
Eq.~(\ref{alau}), the $E_{sym}(\rho)$ depends linearly on the
parameter $x$ at a given density except $\rho_0$ where the symmetry
energy is fixed by construction. As shown in Fig.~\ref{MDIsymE},
adjusting the parameter $x$ in the MDI interaction leads to a broad
range of the density dependence of the nuclear symmetry energy,
similar to those predicted by various microscopic and/or
phenomenological many-body theories.

\subsection{Thermodynamics quantities in neutron stars at $\beta$-equilibrium with
the MDI interaction}\label{therm}

Since we are going to examine astrophysical implications of the
symmetry energy constrained by heavy-ion reactions obtained from
transport model analyses using the MDI interaction, it is useful to
first study several key thermodynamical quantities in neutron star
matter at $\beta$-equilibrium with charge neutrality. It is also
necessary to examine the causality with the MDI interaction.

It is well known that for the $npe\mu$ matter the
$\beta$-equilibrium condition is
\begin{equation}\label{benpemu}
\mu_n-\mu_p=\mu_e=\mu_\mu.
\end{equation}
The appearance of muons requires a sufficiently high chemical
potential of electrons, i.e. $\mu_e>m_\mu$, where $m_\mu$ is the
mass of muons. Eq.~(\ref{benpemu}) together with the charge
neutral condition
\begin{equation} \label{ncnpemu0}
\rho_p=\rho_e+\rho_\mu
\end{equation}
determines the proton fraction $x_p$ as a function of baryon density
in the neutron star matter.

To calculate the core-crust transition density $\rho_t$, we only
need to deal with the $npe$ matter since muons will normally not
appear as the electron chemical potential $\mu_e$ is not high enough
near $\rho_t$ unless one uses an extremely soft symmetry energy. For
the $npe$ matter at $\beta$-equilibrium, one has
\begin{equation}\label{benpe}
\mu_n-\mu_p=\mu_e.
\end{equation}
Then, this identity together with the charge neutral condition
\begin{equation} \label{ncnpemu}
\rho_p=\rho_e
\end{equation}
gives the corresponding $x_p$ as a function of baryon density.

If analytical expressions of the EOS for asymmetric nuclear matter
are known completely as given earlier for the Skyrme and MDI
interactions, the exact $\beta$-equilibrium condition of
Eq.~(\ref{best}) can be used. However, often this is impossible
with many interactions within various models. Instead, the
parabolic approximation of the EOS is usually used. In this case
one has
\begin{equation}\label{bepa}
\mu_e = \mu_\mu \approx 4 (1-2x_p) E_{sym} = 4 \delta E_{sym}.
\end{equation}

\begin{figure}[t!]
\centering
\includegraphics[scale=1.2]{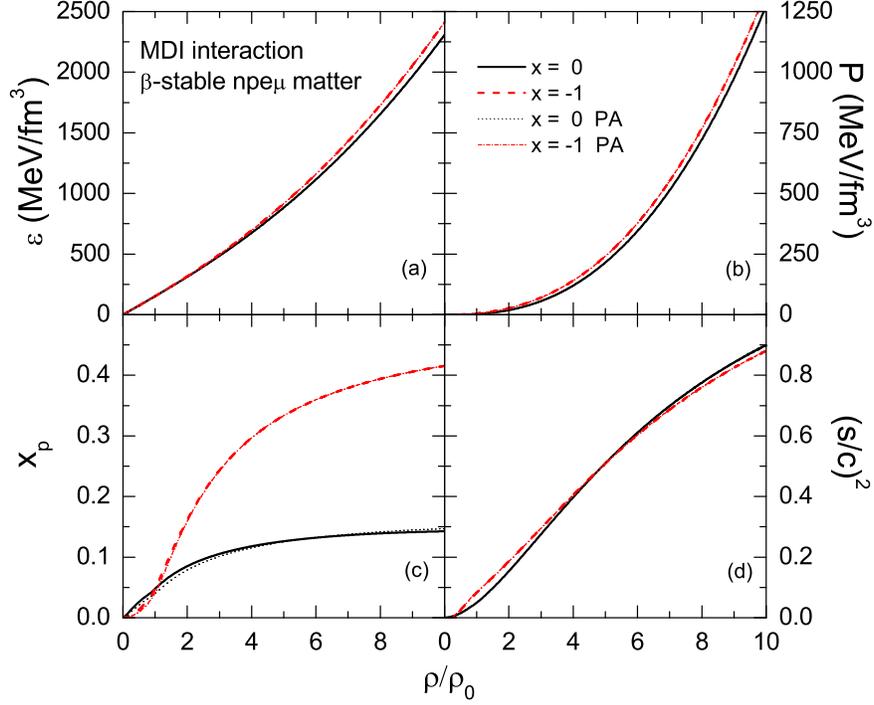}
\caption{{\protect\small (Color online) The density dependence of
the energy density (a), the pressure (b), the proton fraction (c)
and the sound velocity (d) for MDI interaction with $x=0$ and
$x=-1$ for the $npe\mu$ matter at $\beta$-equilibrium. The results
from the full EOS and its parabolic approximation (PA) are
compared.}} \label{MDInpemu}
\end{figure}

Using both the full EOS and its parabolic approximation of the MDI
interaction with $x=0$ and $x=-1$, we have calculated the proton
fraction $x_p$ as a function of density from $0$ to $1.6$ fm$^{-3}$.
The specific values of the $x$ parameter chosen here are consistent
with the constraints extracted from heavy-ion
reactions~\citep{LCK08}. The calculated values of $x_p$ are shown in
the panel (c) of Fig.~\ref{MDInpemu}. Compared to the results with
$x=-1$, the $x_p$ with $x=0$ is larger below the saturation density
and smaller at higher densities. The difference between calculations
using the full EOS and its parabolic approximation is only visible
at low densities for the soft symmetry energy with $x=0$.

We now examine several thermodynamical quantities for the $npe\mu$
matter at $\beta$-equilibrium. The total energy density
$\epsilon(\rho,\delta)$ consists of three parts: the baryon energy
density $\epsilon_b(\rho,\delta)$, the electron energy density
$\epsilon_e(\rho,\delta)$ and the muon energy density
$\epsilon_\mu(\rho,\delta)$
\begin{equation}
\epsilon(\rho,\delta)=\epsilon_b(\rho,\delta)+\epsilon_e(\rho,\delta)+\epsilon_\mu(\rho,\delta),
\end{equation}
where
\begin{equation}\label{VBexact}
\epsilon_b(\rho,\delta) = \rho E_b(\rho,\delta) + \rho m
\end{equation}
with $m$ being the baryon mass and $\rho$ the total baryon density.
The energy density of leptons $\epsilon_l(\rho,\delta)$ is
calculated using the non-interacting Fermi gas model and it can be
expressed as~\citep{Oppen}
\begin{equation}
\epsilon_l(\rho,\delta) = \eta \phi(t),
\end{equation}
with
\begin{equation}
\eta = \frac{m_l c^2}{8 \pi^2 \lambda^3},\notag
\end{equation}
and
\begin{eqnarray}
\lambda &=& \frac{\hbar}{m_l c},
t = \lambda (3 \pi^2 \rho_l)^{1/3}, \notag\\
\phi(t) &=& t \sqrt{1+t^2}(1+2t^2) - \ln(t + \sqrt{1 + t^2}),\notag
\end{eqnarray}
where $m_l$ and $\rho_l$ are the mass and number density of leptons.

Correspondingly, the total pressure $P(\rho,\delta)$ consists of the
contributions from baryons, electrons and muons, i.e.,
\begin{equation}
P(\rho,\delta)=P_b(\rho,\delta)+P_e(\rho,\delta)+P_\mu(\rho,\delta),
\end{equation}
where
\begin{equation}\label{PBexact}
P_b(\rho,\delta)=\mu^\prime_n \rho_n + \mu^\prime_p \rho_p -
\epsilon_b(\rho,\delta),
\end{equation}
and here the chemical potentials should include the rest mass
\begin{equation}
\mu^\prime_n = \mu_n + m,~\mu^\prime_p = \mu_p + m.
\end{equation}
The pressure of leptons is written as
\begin{equation}
P_l(\rho,\delta) = \mu_l \rho_l - \epsilon_l(\rho,\delta),
\end{equation}
where the chemical potential is
\begin{equation}
\mu_l = \sqrt{p_{fl}^2 + m_l^2},
\end{equation}
which is fully determined by the lepton density from
\begin{equation}
p_{fl} = \hbar (3 \pi^2 \rho_l)^{1/3}.
\end{equation}
Then, in this framework the thermodynamical consistency
\begin{equation}\label{thcon}
P= \rho^2 \frac{d \epsilon/\rho }{d \rho}
\end{equation}
is satisfied.

The exact expressions given above can be carried out using the
full EOS. While in some cases, the parabolic approximation is
used. Instead of Eq.~(\ref{VBexact}) and Eq.~(\ref{PBexact}),
within the parabolic approximation one has
\begin{equation}\label{VBPA}
\epsilon_b(\rho,\delta) = \rho [E_0(\rho) + E_{sym}(\rho) \delta^2]
+ \rho m,
\end{equation}
\begin{equation}\label{PBPA}
P_b(\rho,\delta)=\rho^2(E^\prime_0(\rho) + E^\prime_{sym} (\rho)
\delta^2).
\end{equation}
This approximation still satisfy the thermodynamical consistency
(Eq.~(\ref{thcon})).

The density dependence of the total energy density and pressure
are shown in Panel (a) and Panel (b) of Fig.~\ref{MDInpemu},
respectively. The difference between calculations using the full
EOS and its PA is essentially invisible. The stiffer (e.g., x=-1)
the symmetry energy is, the larger the total energy and pressure
are as one expects.

The causality requires that the sound speed $s$ in nuclear matter
remains smaller than the speed of light in vacuum $c$, i.e.,
\begin{equation}
\frac{s}{c} = \sqrt{\frac{\partial P}{\partial \epsilon}} < 1.
\end{equation}
In Panel (d) of Fig.~\ref{MDInpemu} we examine the speed of sound
for the MDI interaction with $x=0$ and $x=-1$. It is seen that the
causality is satisfied in the whole density range considered.

\section{Key equations for describing the structure of neutron stars}
\label{nstarth}

For completeness, we quote here from the general literature, see,
e.g., ref.~\citep{Morrison}, some key equations to be used later in
our studies of neutron star structure.  For slowly-rotating neutron
stars where the spherical symmetry is conserved approximately, the
moment of inertia is
\begin{equation}\label{I}
I=(\frac{\partial J}{\partial \Omega })_{\Omega =0}=\frac J\Omega,
\end{equation}
where $\Omega $ is the angular velocity measured in a far-away
inertial system and $J$ is the angular momentum. In the
slow-rotation limit in spherical polar coordinates the metric can be
written as ($G=c=1$)
\begin{eqnarray}\label{metric}
(ds)^2 &=& -e^{\nu}(dt)^2+(1-\frac{2m_g}{r})^{-1}(dr)^2 \notag\\
&-&2\omega r^2 \sin^2\theta dt d\phi + r^2(d \theta^2 + \sin^2
\theta d \phi^2),
\end{eqnarray}
where $\omega (r)\equiv \frac{d \phi}{dt}$ is the angular velocity
of the local slow-rotation system (measured in a far-away inertial
system), and $m_g(r)$ is the neutron star gravitational mass inside
a radius $r$
\begin{equation}\label{m}
\frac{dm_g(r)}{dr} = 4\pi r^2\epsilon (r),
\end{equation}
with $\epsilon (r)$ being the energy density. Defining
\begin{equation}
\overline{\omega} =\Omega -\omega,  \label{omegabar}
\end{equation}
then from the metric of the geometry outside a slow-rotation star
one can get
\begin{equation}
r^{-4}\frac d{dr}[r^4j(r)\frac{d\overline{\omega }}{dr}]+4r^{-1}\frac{dj}{dr}%
\overline{\omega }(r) =0,\label{eomegabar}
\end{equation}
where
\begin{equation}
j(r)=\exp [-\frac 12(\lambda(r) +\nu(r) )],
\end{equation}
with
\begin{equation}
e^{\lambda(r)} = (1-\frac{2m_g}{r})^{-1},
\end{equation}
and
\begin{equation}\label{nu}
\frac{d \nu}{d r} = -\frac{2}{\epsilon+P} \frac{d P}{d r}.
\end{equation}
The pressure $P(r)$ is obtained from the famous
Tolman-Oppenheimer-Volkoff (TOV) equation, i.e.,
\begin{equation}\label{tov}
\frac{d P}{d r} = -(\epsilon+P)\frac{m_g+4 \pi r^3 P}{r(r-2m_g)}.
\end{equation}

For convenience, an additional function $\eta(r)$ can be defined as
\begin{equation}
\eta(r) = \frac{d\overline{\omega }}{dr},\label{eta}
\end{equation}
and the boundary conditions at the central of the star are
\begin{equation}\label{bomega}
\overline{\omega }(0)=const,
\end{equation}
and the constant is chosen so that
\begin{equation}\label{beta}
\eta(0)=0.
\end{equation}
Outside the star we should have
\begin{equation}
e^\nu = 1-\frac{2M}{r}
\end{equation}
and
\begin{equation}
\omega=\frac{2J}{r^3},
\end{equation}
where
\begin{equation}
J=\frac 16 R^4 \eta(R) \label{J}
\end{equation}
and the $M$ and $R$ are total gravitational mass and the total
radius of the neutron star, respectively. To make the variables
continuous at the surface of the star, we have the boundary
conditions
\begin{equation}\label{bR}
\overline{\omega }(R) = \Omega - \eta(R)\frac{R}{3},\;\;\nu(R) =
\ln(1-2M/R).
\end{equation}
In this framework, after solving the differential equations
Eq.~(\ref{m}), (\ref{eomegabar}), (\ref{tov}) and (\ref{nu}), the
mass, radius and moment of inertia can be calculated. Following the
standard procedure, we integrate out the TOV and other differential
equations equation starting from the center to the surface where the
pressure vanishes, i.e., $ P(R)=0.$ The latter defines the total
radius $R$ of the neutron star. Then the total gravitational mass of
the neutron star is obtained from integrating Eq.~(\ref{m}) as
\begin{equation}
M\equiv m_g(R)=4\pi \int_0^Rdrr^2\epsilon (r).
\end{equation}
The total moment of inertia of the neutron star is obtained
similarly. By integrating only to the transition density $\rho_t$,
one can obtain the radius and mass of the core. The thickness, mass
and moment of inertia of the crust can be obtained from taking the
differences between values for the whole and the core of neutron
stars.

\section{Results and discussions}
\label{results}

In the following, we present and discuss results of our calculations
on the transition density and pressure at the inner edge and several
global properties of neutron stars. Applying formalisms outlined in
the previous sections, we illustrate numerically and discuss several
issues including (a) relationships among the mechanical, chemical
and total instability boundaries in asymmetric nuclear matter, and
their relevance for locating the core-crust transition density in
the $npe$ matter at $\beta$-equilibrium; (b) the difference between
the core-crust transition densities obtained using the dynamical and
thermodynamical methods using the same interactions; (c)
understanding the difference between the core-crust transition
densities obtained with the full EOS and its parabolic approximation
using the same methods and interactions; (d) the systematics of the
transition density by varying the stiffness of the symmetry energy;
(e) limits on the transition density using the symmetry energy
constrained by heavy-ion experiments; (f) relationship between the
transition density and the size of neutron-skin in $^{208}$Pb; (g)
systematics and constraints on the transition pressure at the
core-crust boundary. We will then study several global properties of
neutron stars including the mass, radius, and the moment of inertia
as well as their crustal fractions. The focus will be on effects of
the density dependence of the symmetry energy on these observables.
We will also check the inner crust EOS dependence of properties of
neutron stars.

\subsection{Instabilities in neutron-rich matter and the core-crust transition density
in neutron stars at $\beta$-equilibrium}\label{rhot}

\begin{figure}[t!]
\centering
\includegraphics[scale=1.2]{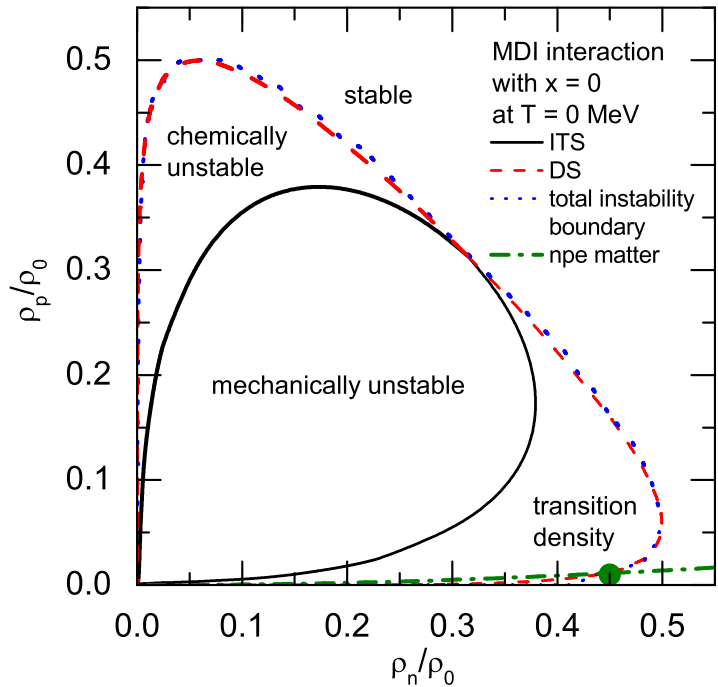}
\caption{{\protect\small (Color online) The mechanical, chemical
and total instability boundaries shown in the $\rho_n \sim \rho_p$
plane using the MDI interaction with $x=0$ at zero temperature.
The $\rho_n$ vs. $\rho_p$ for the $npe$ matter at
$\beta$-equilibrium is shown as the dash-dot line. The core-crust
transition density from the thermodynamical method is indicated
with the filled dot.}} \label{mdiinstability}
\end{figure}

In the subsection \ref{relation} we studied analytically the
relationship between the dynamical and thermodynamical methods. To
appreciate the relationship more clearly and quantitatively, we
present here a numerical example in the $\rho_p$ vs. $\rho_n$ plane.
First, it is important to recognize that the right-hand side of
Eq.~(\ref{relation2}) just determines the thermodynamical
instability of asymmetric nuclear matter. Shown in
Fig.~\ref{mdiinstability} are the boundaries of the mechanical (also
known as the isothermal spinodal (ITS)), chemical (also known as the
diffusive spinodal (DS)) and total instabilities without requiring
the $\beta$-equilibrium and charge neutrality using the MDI
interaction with $x=0$ at zero temperature. A similar figure has
been shown in our previous paper~\citep{Xu07b} but in the
$\rho\sim\delta$ or $P\sim\rho$ plane at finite temperatures. Inside
the ITS curve the system is mechanically unstable, while between the
ITS curve and the DS curve the system is chemically unstable. The
total instability is identified by the condition
\begin{equation}
\frac{\partial \mu_n}{\partial \rho_n}\frac{\partial \mu_p}{\partial
\rho_p}-\left(\frac{\partial \mu_n}{\partial \rho_p}\right)^2<0.
\end{equation}
It is seen clearly that the total instability region obtained using
the above condition covers the region of both mechanical and
chemical instabilities. This observation is consistent with the
earlier finding by Margueron and Chomaz~\citep{Chomaz03}. In the
$npe$ matter when the $\beta$-equilibrium and charge neutrality
conditions are imposed, the $\rho_n$ and $\rho_p$ are correlated
with each other. For the MDI interaction, this correlation can be
obtained from the $x_p$ versus $\rho$ curves shown in the window (c)
of Fig.~\ref{MDInpemu}. With $x=0$, this correlation is shown with
the dash-dot line in Fig.~\ref{mdiinstability}. The cross point of
this line and the boundary of total instability corresponds to the
core-crust transition density in the $npe$ matter within the
thermodynamical approach. The density gradient term and the Coulomb
interaction generally reduce slightly the instability region, thus
the dynamical method normally leads to a slightly lower transition
density.

It is necessary to note here that the onset of instabilities has
been associated with the so-called liquid-gas phase transition in
nuclear matter~\citep{Sie83}. An experimental manifestation of the
liquid-gas phase transition is the well-known multifragmentation
phenomenon in heavy-ion collisions, see, e.g.,
refs.~\citep{Cho04,Das05} for a recent review. It is seen from
Fig.~\ref{mdiinstability} that both the dynamical and
thermodynamical models give a liquid-gas phase transition density of
about $0.63\rho_0$ for symmetric (i.e., $\rho_n=\rho_p$) nuclear
matter (SNM) at $T=0$. While the latter depends slightly on the
interaction used, it is consistent with previous calculations on the
boundary of mechanical instability in cold SNM, see, e.g.,
refs.~\citep{Mul95,LiKo97}. We thus conclude that both the dynamical
and thermodynamical methods give the right asymptotical value for
the transition density when one goes from the $npe$ matter to the
symmetric nuclear matter at T=0.

\subsection{Constraining the core-crust transition density in neutron stars}

\begin{figure}[t!]
\centering
\includegraphics[scale=1.2]{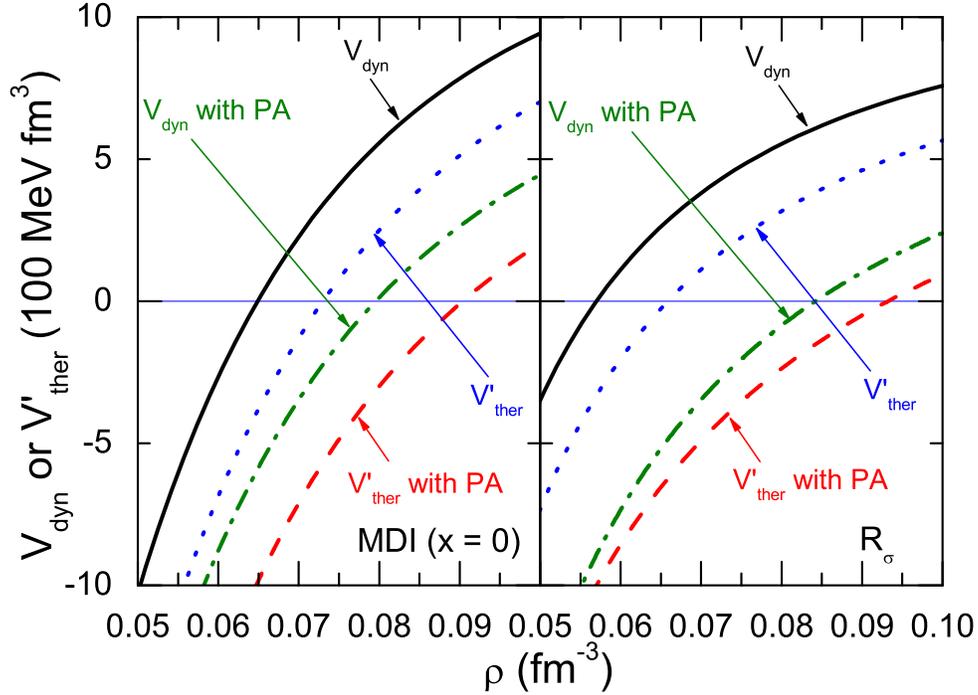}
\caption{{\protect\small (Color online) The density dependence of
$V$ for the MDI interaction with $x=0$ (left window) and the Skyrme
force R$_\sigma$ (right window) using both the dynamical and
thermodynamical methods with the full EOS and its parabolic
approximation (PA).}} \label{rsigmaV}
\end{figure}

We now turn to the numerical calculations and comparisons of the
core-crust transition densities within both the dynamical and
thermodynamical methods using the full EOS and its PA with the MDI
and Skyrme interactions. The transition density can be directly
obtained by carrying out the analyses in the $\rho$ vs. $x_p$ plane.
We stress again that in principle the transition density should be
calculated from Eq.~(\ref{Vdynmin}) or Eq.~(\ref{Vther}), and the
$\beta$-equilibrium condition should be expressed as
Eq.~(\ref{best}). While in practical calculations, the parabolic
approximation has often been used in determining the
$\beta$-equilibrium condition using Eq.~(\ref{bepa}) and in
evaluating the $V_{ther}$ using Eq.~(\ref{Vtherpa}). To first
evaluate effects of the PA, we show in Fig.~\ref{rsigmaV} the
density dependence of $V_{dyn}$ and $V^\prime_{ther}$ using the MDI
interaction with $x=0$ and the Skyrme force R$_\sigma$ within both
the dynamical and thermodynamical methods with the full EOS and its
PA. Here, we have defined
\begin{equation}
V^\prime_{ther}=V_{ther} \frac{\partial^2 E_b}{\partial x_p^2}
/\left(\rho^2 \frac{\partial \mu_n}{\partial \rho_n}\right)
\end{equation}
and it should be noted that $V^\prime_{ther}$ has the same vanishing
point as the $V_{ther}$ and the same dimension as the $V_{dyn}$. For
the MDI interaction with $x=0$ the transition densities using the
full EOS within the dynamical and thermodynamical method are $0.065$
fm$^{-3}$ and $0.073$ fm$^{-3}$, respectively. While the
corresponding results using the PA are $0.080$ fm$^{-3}$ and $0.090$
fm$^{-3}$, respectively. For the Skyrme force R$_\sigma$ the
transition densities are $0.057$ fm$^{-3}$ and $0.066$ fm$^{-3}$
using the full EOS, while the corresponding values with the PA are
$0.084$ fm$^{-3}$ and $0.093$ fm$^{-3}$, by using the dynamical and
thermodynamical method, respectively. Thus, the transition densities
are generally lower with the dynamical method as we mentioned
earlier, as the density gradient term and the Coulomb interaction
make the system more stable. However, the PA significantly lifts the
transition density regardless of the approach used. In fact, the
difference between calculations using the full EOS and its PA is
much larger than that caused by using the two different methods.

\begin{figure}[t!]
\centering
\includegraphics[scale=1.2]{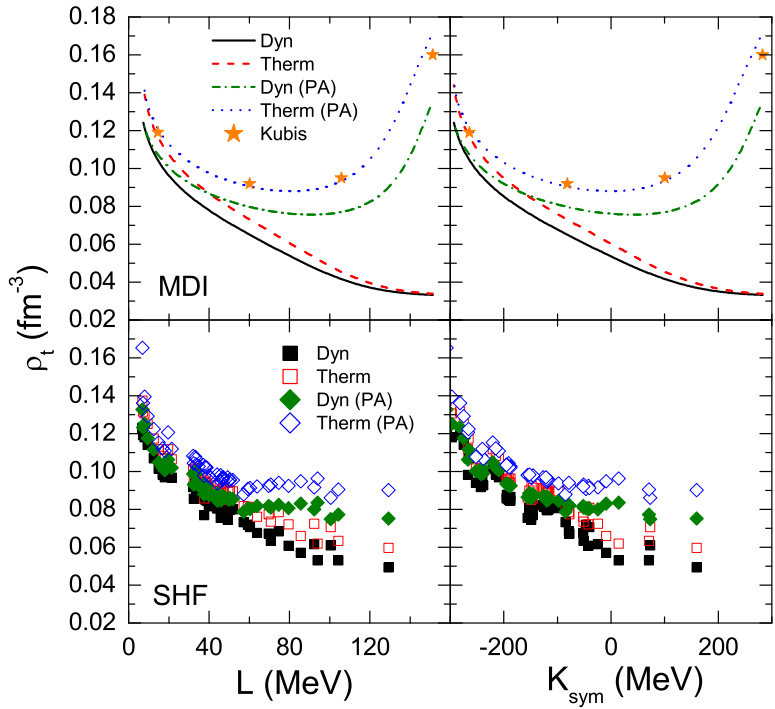}
\caption{{\protect\small (Color online) The transition density as
a function of $L$ (left panel) and $K_{sym}$ (right panel) by using
both the dynamical and thermodynamical methods with the full EOS
and its parabolic approximation. The MDI (upper windows) and
Skyrme interactions (lower windows) are used.}} \label{rhotLK}
\end{figure}

As we have mentioned in the introduction, it is well known that the
transition density depends sensitively on the $E_{sym}(\rho)$. Many
interesting studies, see, e.g., refs.~\citep{Oya07,Lat07,Dou00},
have been carried out using various $E_{sym}(\rho)$. In the
following, we present and compare the systematics of the transition
density using the MDI interaction with the varying $x$ parameter and
$51$ Skyrme forces. Since the density dependence of the symmetry
energy can be well characterized by the $L$ and $K_{sym}$
parameters, we examine the $\rho_t$ as a function of $L$ and
$K_{sym}$ in Fig.~\ref{rhotLK}. Shown in the left panels are the
$\rho_t$ as a function of $L$ by using both the dynamical and
thermodynamical methods with the full EOS and its PA. The same
quantities are shown as a function of $K_{sym}$ in the right panels.
It is interesting to see that both the dynamical and thermodynamical
methods give very similar results with the former giving slightly
smaller $\rho _{t}$ than the later (the difference is actually less
than $0.01$ fm$^{-3}$) and this is due to the fact that the former
includes the density gradient and Coulomb terms which make the
system more stable and lower the transition density. The small
difference between the two methods implies that the effects of
density gradient terms and Coulomb term are unimportant in
determining the $\rho _{t}$. In addition, it is also interesting to
see that the transition density decreases almost linearly with the
increasing $L$ especially in the calculations with the full EOS.
This observation is consistent with that found recently by Oyamatsu
et al.~\citep{Oya07}. We note here that there are some interactions
with larger $x$ values in the MDI interaction giving negative and/or
very small values for the $L$ parameter. These interactions with
negative and/or very small values for the $L$ parameter, however,
lead to neutron-skins in $^{208}$Pb inconsistent with the existing
data~\citep{Ste05b,Che05b}. Since they are still somewhat
theoretically interesting, we have thus also examined the possible
transition density with these interactions. We find that for the
interaction parameters with $L<7$ MeV in the MDI interaction
($x>1.17$), the transition density does not exist and the $npe$
matter is always unstable. This is due to the fact that the symmetry
energy is so soft that the $\frac{\partial \mu_n}{\partial \rho_n}$
is always negative while the $\frac{\partial \mu_p}{\partial
\rho_p}$ is always positive at low densities, and thus the stability
condition $\frac{\partial \mu_n}{\partial \rho_n}\frac{\partial
\mu_p}{\partial \rho_p}-\left(\frac{\partial \mu_n}{\partial
\rho_p}\right)^2>0$ can never be satisfied.

It is clear from all existing calculations that the $\rho_t$ is
sensitive to the density dependence of the nuclear symmetry energy.
The latter can be well characterized by the slope $L$ and the
curvature $K_{sym}$. Naturally, there are some correlations between
the $L$ and $K_{sym}$ determined by the interaction energy density
functional used. For the MDI interaction, the $L$ and $K_{sym}$ both
change linearly with the parameter $x$. Therefore they are linearly
correlated. Similarly, the $L$ and $K_{sym}$ also correlated within
the SHF model. It is therefore not surprising that the variation of
$\rho_t$ with $K_{sym}$ is very similar to that with $L$, as shown
in the right panels of Fig.~\ref{rhotLK}.

We now apply the experimentally constrained $L$ to the $\rho_t-L$
correlation obtained using the full EOS within the dynamical method
in constraining the $\rho _{t}$. In our earlier transport model
studies of the isospin diffusion data in heavy-ion
reactions\citep{Che05a,LiBA05c}, the complete MDI interaction was
used. The extracted $L$ value is $88\pm25$ MeV if one defines the
$E_{sym}$ using the PA in Eq.~(\ref{Esym-pa}) or $86\pm25$ MeV if
one uses the exact expression of $E_{sym}$(Eq.~(\ref{esymmdi}))
corresponding to the full MDI EOS. Using the latter in comparison
with the full MDI results shown in Fig.~\ref{highorder}, we conclude
that the transition density is between $0.040$ fm$^{-3}$ and $0.065$
fm$^{-3}$. This constrained range is significantly below the
fiducial value of $\rho_t\approx 0.08$fm$^{-3}$ often used in the
literature and the estimate of $0.5<\rho_t/\rho_0<0.7$ made in
ref.~\citep{Lat07} within the thermodynamical approach using the
parabolic approximation of the EOS. This difference is
understandable as we shall explain in detail below.

\subsection{Understanding effects of the parabolic approximation of the EOS on the
core-crust transition density in neutron stars}

\begin{figure}[t!]
\centering
\includegraphics[scale=1.2]{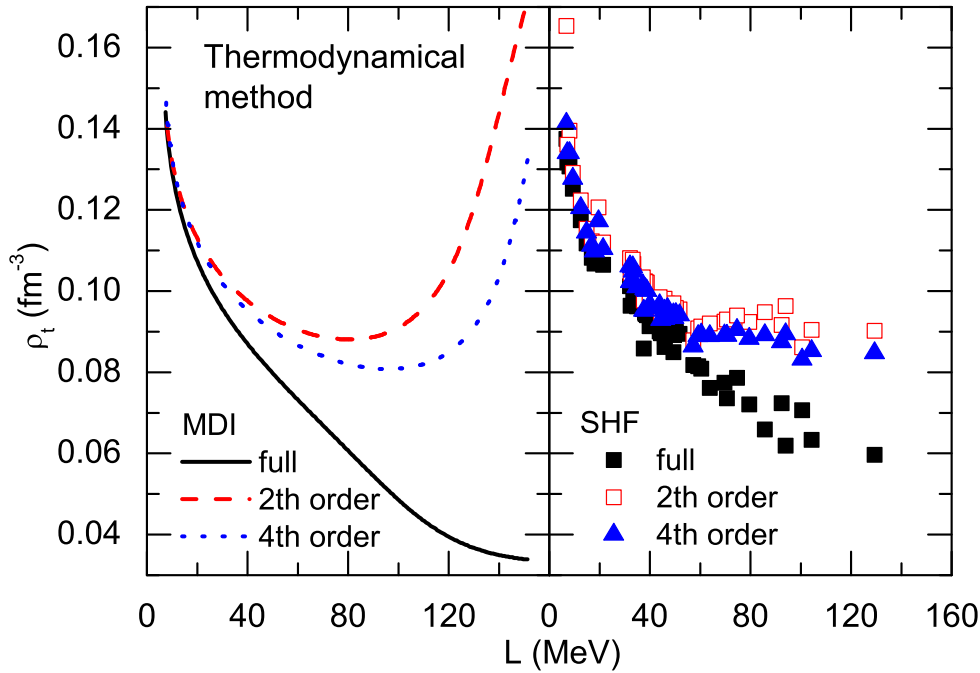}
\caption{{\protect\small (Color online) The relation between the
transition density and $L$ by using the thermodynamical method with the
MDI (left panel) and Skyrme (right panel) interactions. For both kinds of
interactions, up to the 2nd and 4th orders in isospin asymmetry are
used in the expansions of the corresponding full EOS.}} \label{highorder}
\end{figure}

It is also seen from Fig.\ \ref{rhotLK} that except at very small
values of $L$ and $K_{sym}$, there is a big difference between
results obtained using the full EOS and its parabolic approximation
within both the dynamical and thermodynamical methods. Especially at
high values of $L$ and $K_{sym}$, the $\rho_t$ from the PA increases
while the one from the full EOS continuously decreases. We also
notice that Kubis's calculations~\citep{Kub07} coincide with our
results using the MDI interaction within the thermodynamical method
with the PA. Why is the transition density so sensitive to whether
one used the PA or not? To answer this question, we first notice
that both the first and second derivatives of the EOS are involved
in the stability conditions. The EOS can be expanded according to
$E_b(\rho,x_p)$ up to the fourth order term of $(1-2x_p)$ according
to
\begin{eqnarray}
E_b(\rho,x_p) &=& E_0(\rho) + E_{sym}(\rho)(1-2x_p)^2 + E_4(\rho)(1-2x_p)^4 \notag\\
&+& O(1-2x_p)^6.
\end{eqnarray}
Only even order terms of $(1-2x_p)$ appear as the strong interaction
is assumed to be symmetric for exchanging neutrons with protons. The
first and second order derivatives of the energy with respect to
$x_p$ are, respectively,
\begin{equation}
\frac{\partial E_b}{\partial x_p} = -4 E_{sym}(\rho)(1-2x_p) - 8
E_4(\rho) (1-2x_p)^3 + O(1-2x_p)^5.
\end{equation}
\begin{equation}
\frac{\partial^2 E_b}{\partial x_p^2} = 8 E_{sym}(\rho) + 48
E_4(\rho) (1-2x_p)^2 + O(1-2x_p)^4.
\end{equation}
At $\beta$-equilibrium the $npe$ matter is usually highly neutron
rich, so the $(1-2x_p)$ is not far from $1$. Thus the higher order
terms in $(1-2x_p)$ are normally not negligible. Moreover, although
the coefficient $E_4$ is normally smaller than the $E_{sym}$, the
contribution to the $\frac{\partial E_b}{\partial x_p}$ and the
$\frac{\partial^2 E_b}{\partial x_p^2}$ from the $E_4$ term gains a
factor of $2$ and $6$, respectively, compared to that from the
$E_{sym}$ term. Thus, mathematically one expects the $E_4$ term to
play an important role in locating the stability boundaries in
asymmetric nuclear matter and the core-crust transition density in
neutron stars. It is also easy to understand why the effect is
stronger with stiffer symmetry energy functionals. At sub-saturation
densities near the $\rho_t$, the proton fraction $x_p$ is lower with
the stiffer symmetry energy. It is the opposite at supra-saturation
densities. A numerical example can be found in the window (c) of
Fig.~\ref{MDInpemu} for the MDI interaction with $x=0$ (softer) and
$x=-1$ (stiffer). It is seen that with the stiffer symmetry of
$x=-1$, near the transition density $(1-2x_p)$ is indeed larger than
that with a softer symmetry energy of $x=0$. Therefore, a larger
error will be introduced in calculating the $\rho_t$ using the
parabolic approximation with stiffer symmetry energy functionals. To
be more quantitative, we compare in Fig.~\ref{highorder} the
$\rho_t$ as a function of $L$ obtained within the thermodynamical
method using the full EOS with those obtained by expanding the EOS
to the second and 4th orders in $(1-2x_p)$. The left window is for
calculations with the MDI interaction, and the right one with the
$51$ Skyrme forces. We notice that the convergence speed is very
slow, and not only the fourth order term but also the sixth, eighth
or even further higher order terms should be considered (For the
Skyrme forces, we noted the calculations up to the 8th order
approximation still leads to a significant error compared to the
full EOS). We also notice here that the EOS of asymmetric nuclear
matter always contains the higher-order terms in isospin asymmetry
at least due to the kinetic contribution. Moreover, if we use the
Eq.~(\ref{Esym-pa}) instead of Eq.~(\ref{Esym-d}) in calculating the
symmetry energy and then reconstruct the EOS as $E(\rho ,\delta
)=E(\rho ,\delta =0)+[E(\rho ,\delta=1)-E(\rho ,\delta =0)]\delta
^{2}+O(\delta^{4})$ as in the parabolic approximation, almost the
same transition densities are obtained as the second order
approximation shown in Fig.\ \ref{highorder}. Our results thus
indicate that one may introduce a huge error by assuming {\it a
priori} that the EOS is parabolic for a given interaction in
calculating the $\rho _{t}$. It is thus clear that the correct
transition density can hardly be obtained without knowing the exact
expression of $E_b(\rho,x_p)$ for a given interaction.
Interestingly, these features agree with the early finding
\citep{Arp72} that the $\rho _{t}$ is very sensitive to the fine
details of the nuclear EOS.

\subsection{Correlation between the core-crust transition density
in neutron stars and the size of neutron-skin in $^{208}$Pb}

\begin{figure}[t!]
\centering
\includegraphics[scale=1.2]{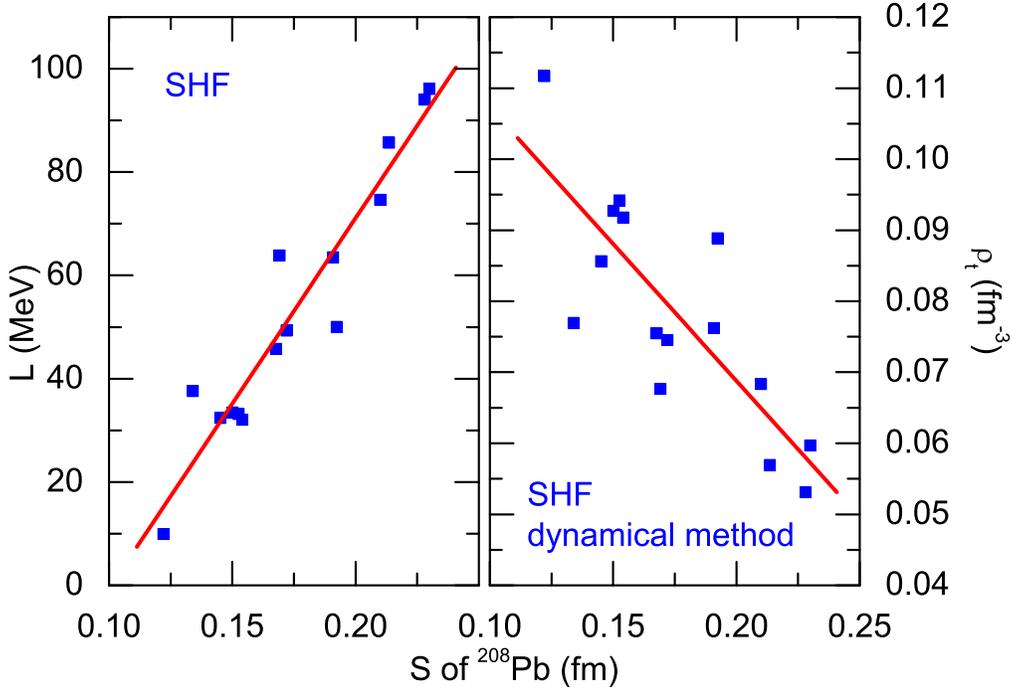}
\caption{{\protect\small (Color online) The slope parameter $L$ and
the transition density $\rho_t$ as a function of the neutron skin
thickness $S$ of $^{208}$Pb using the dynamical method with the full
EOS with the Skyrme interactions. The solid lines indicate the
linear fits.}} \label{SrhotL}
\end{figure}
It is also well known that the sizes of neutron skins in heavy
nuclei are sensitive to the symmetry energy at subsaturation
densities, see, e.g.,
refs.~\citep{Brown00,Hor01,Fur02,Die03,Ste05,Tod05,Che05b}. However,
available data of neutron-skin thickness obtained using hadronic
probes are not accurate enough yet to constrain tightly the symmetry
energy. Interestingly, the parity radius experiment (PREX) at the
Jefferson Laboratory aiming to measure the neutron radius of
$^{208}$Pb via parity violating electron scattering (Jefferson
Laboratory Experiment E-00-003)~\citep{Horowitz:2001} hopefully will
provide much more precise data in the near future. It can then
potentially constrain the symmetry energy at low densities and thus
the core-crust transition density more stringently. It has been
shown by many authors and also in our previous work~\citep{Che05b}
that the neutron skin thickness $S$ increases linearly with $L$.
Given the fact that the transition density $\rho_t$ decreases almost
linearly with the increasing $L$ as shown above, it is interesting
to examine the correlation between the $S$ and $\rho_t$. Such kind
of study was first carried out in ref.~\citep{Hor01} using the RMF
EOS and the $\rho_t$ calculated within the RPA approach.

Shown in Fig.~\ref{SrhotL} are the $\rho_t$ and $L$ versus the
neutron skin thickness $S$ of $^{208}$Pb obtained by using the
dynamical method with the full SHF EOS. As known before, the neutron
skin thickness increases linearly with the increasing
$L$~\citep{Che05b}. Moreover, the transition density shows a
decreasing trend with the increasing neutron skin thickness. As the
$\rho_0$ and $E_{sym}(\rho_0)$ are different for the various sets of
Skyrme forces, the data points do not show a very strong linear
correlation. However, the tendency is clear. This trend is
consistent with the RPA calculations using the RMF EOS's by Horowitz
et al.~\citep{Hor01,Hor03}.

\subsection{The pressure at the inner edge of neutron star crust}\label{Pt}

\begin{figure}[t!]
\centering
\includegraphics[scale=1.2]{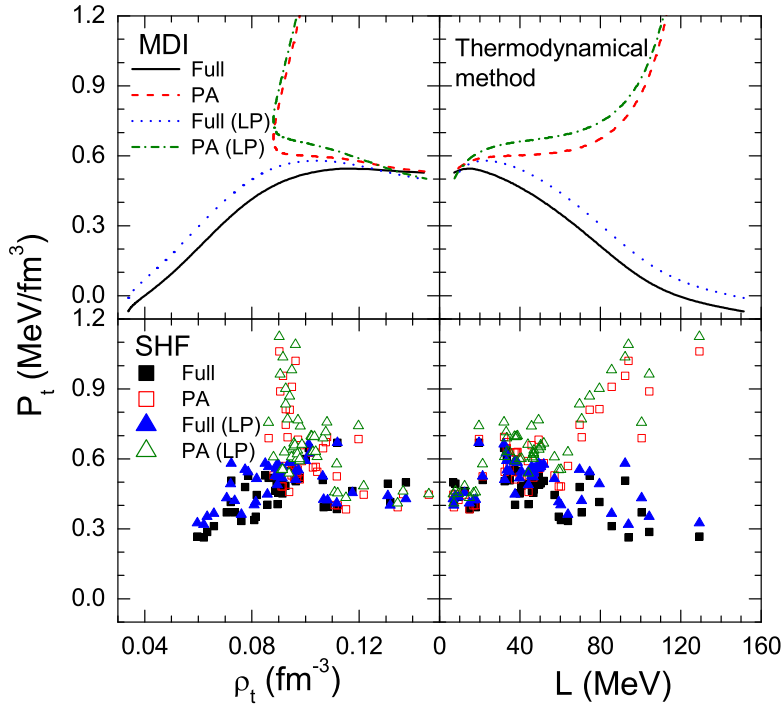}
\caption{{\protect\small (Color online) The transition pressure
$P_t$ as a function of $\rho_t$ and $L$ within the thermodynamical
method with the full EOS and its parabolic approximation using the
MDI (upper windows) and Skyrme (lower windows) interactions.}}
\label{PtrhotLther}
\end{figure}

\begin{figure}[t!]
\centering
\includegraphics[scale=1.2]{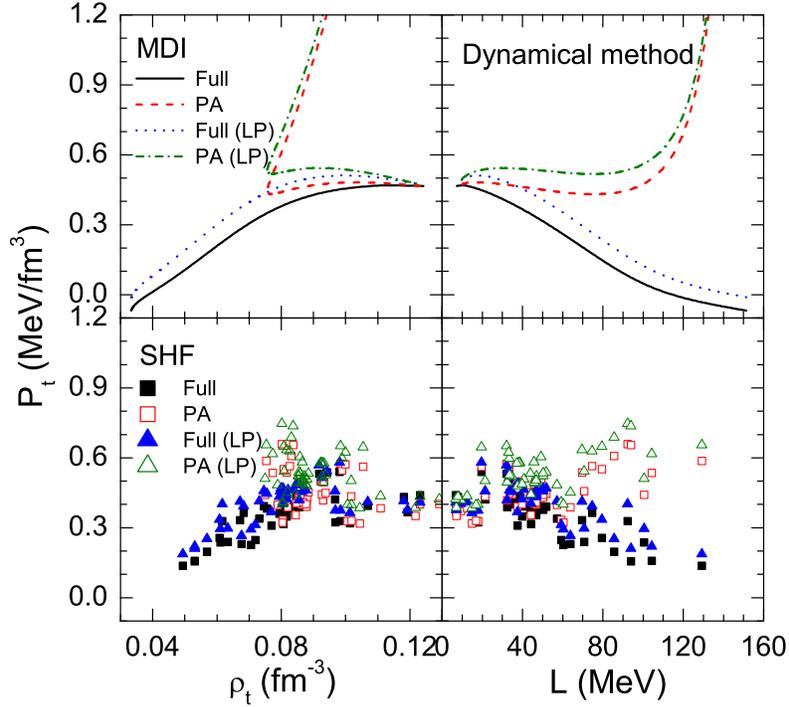}
\caption{{\protect\small (Color online) Same as
Fig.~\ref{PtrhotLther} but within the dynamical method.}}
\label{PtrhotLdyn}
\end{figure}
The pressure at the inner edge is an important quantity related
directly with the crustal fraction of the moment of inertia which
can be measurable indirectly from observations of pulsar
glitches~\citep{Lat07}. In principle, having determined the
transition density it is straightforward to calculate the
corresponding pressure using the formalisms outlined in the
subsection~\ref{therm}. Before presenting the numerical results, it
is very instructive to quote the analytical estimation obtained by
Lattimer and Prakash~\citep{Lat07} for the transition pressure
\begin{eqnarray}\label{lp}
P_t &=&
\frac{K_0}{9}\frac{\rho_t^2}{\rho_0}\left(\frac{\rho_t}{\rho_0}-1\right)
+\rho_t\delta_t
\left[\frac{1-\delta_t}{2}E_{sym}(\rho_t)+\left(\rho\frac{d
E_{sym}(\rho)}{d \rho}\right)_{\rho_t}\delta_t\right],
\end{eqnarray}
where $K_0$ is the incompressibility of SNM at $\rho_0$ and
$\delta_t$ is the isospin asymmetry at $\rho_t$. Besides the
implicit dependence on the symmetry energy through the $\rho_t$ and
$\delta_t$, the $P_t$ also depends explicitly on the value and slope
of the $E_{sym}(\rho)$ at $\rho_t$. Thus the $P_t$ depends very
sensitively on the $E_{sym}(\rho)$. Noticing that the Eq.~(\ref{lp})
is an estimate using the thermodynamical method with the PA, it is
interesting to compare numerically its predictions with the results
obtained according to the original formalisms in the
subsection~\ref{therm} using both the full EOS and its PA within the
thermodynamical and dynamical method, separately.

\begin{figure}[t!]
\centering
\includegraphics[scale=1.2]{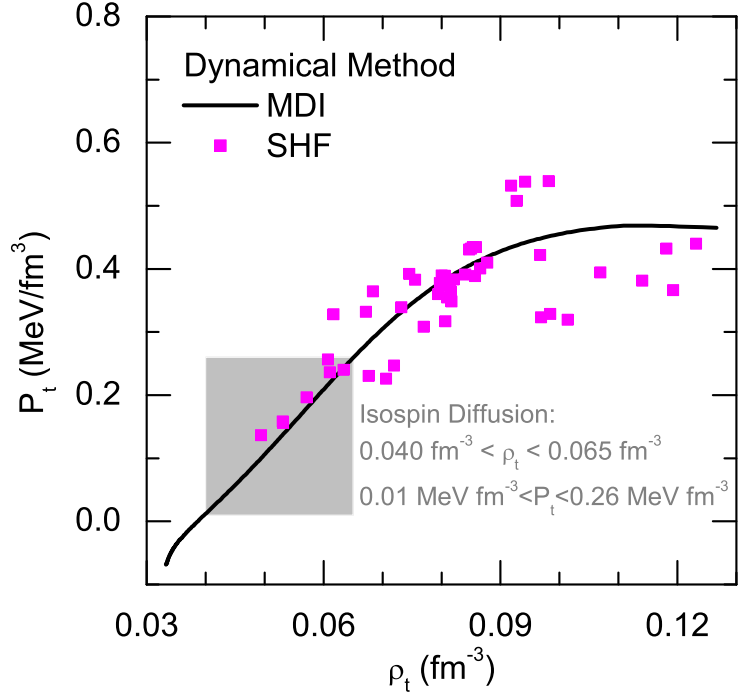}
\caption{(Color online) $P_t$ as a function of $\rho_t$ by using
dynamical method without parabolic approximation for both MDI
interaction and SHF calculations. The shaded band represent the
constraint from the isospin diffusion data.} \label{PtRhotDyn}
\end{figure}

In Fig.~\ref{PtrhotLther} we show the $P_t$ as a function of
$\rho_t$ (left windows) and $L$ (right windows) by using the
thermodynamical method with and without the parabolic approximation.
The same quantities with the dynamical method are shown in
Fig.~\ref{PtrhotLdyn}. Both the MDI (upper windows) and Skyrme
(lower windows) interactions are used. The results from
Eq.~(\ref{lp}) using the $\rho_t$ and $E_{sym}$ corresponding to the
full EOS and its PA are also shown for comparisons. It is
interesting to see that the Eq.~(\ref{lp}) predicts qualitatively
the same but quantitatively slightly higher values compared to the
original expressions for the pressure with or without the PA for
both the thermodynamical and dynamical methods even though this
formula was derived from the thermodynamical method using the PA.
This observation is consistent with the results shown in the window
(b) of Fig.\ \ref{MDInpemu}, namely, the direct effect of using the
full EOS or its PA on the pressure is small although the PA may
affect strongly the transition pressure $P_t$ by changing the
transition density $\rho_t$. The $P_t$ essentially increases with
the increasing $\rho_t$ in calculations using the full EOS, but a
complex relation between the $P_t$ and $\rho_t$ is obtained using
the PA. The observed large difference in $P_t$ is due to the strong
PA effect on the $\rho_t$. Moreover, the latter does not vary
monotonically with $L$ for the PA as shown in Fig.~\ref{rhotLK}.
Thus the PA of the EOS leads to a very different $P_t$ compared to
the calculations with the full EOS especially for the stiffer
symmetry energy functionals.

It is also interesting to examine the range of $P_t$ corresponding
to the $\rho_t$ and $L$ constrained by the heavy-ion reaction data.
In Fig.~\ref{PtRhotDyn}, we show the $P_t$ as a function of $\rho_t$
by using the dynamical method and the full EOS for both the MDI
(solid line) and the Skyrme (filled squares) calculations. It is
interesting to see that the MDI and Skyrme interactions give
generally quite consistent results. Corresponding to the $\rho_t$
constrained in between 0.040 fm$^{-3}$ and 0.065 fm$^{-3}$, the
$P_{t}$ is limited between $0.01$ MeV/fm$^{3}$ and $0.26$
MeV/fm$^{3}$ with the MDI interaction as indicated by the shaded
area, which is significantly less than the fiducial value of
$P_t\approx 0.65$ MeV/fm$^{3}$ often used in the literature
~\citep{Lat07}. As pointed out in a recent work by Avancini et
al~\citep{Ava08b}, the value of $P_t\approx 0.65$ MeV/fm$^{3}$ may
be too large for most mean-field calculations without the PA. We
notice here that among the $51$ Skyrme interactions listed in
Tables~\ref{SHFtab1} and ~\ref{SHFtab2}, the following $7$
interactions, i.e., the SkMP, SKO, R$_\sigma$, G$_\sigma$, SkI2,
SkI3, and SkI5, are consistent with the constraints from heavy-ion
reactions.

\begin{figure}[t!]
\centering
\includegraphics[scale=1.2]{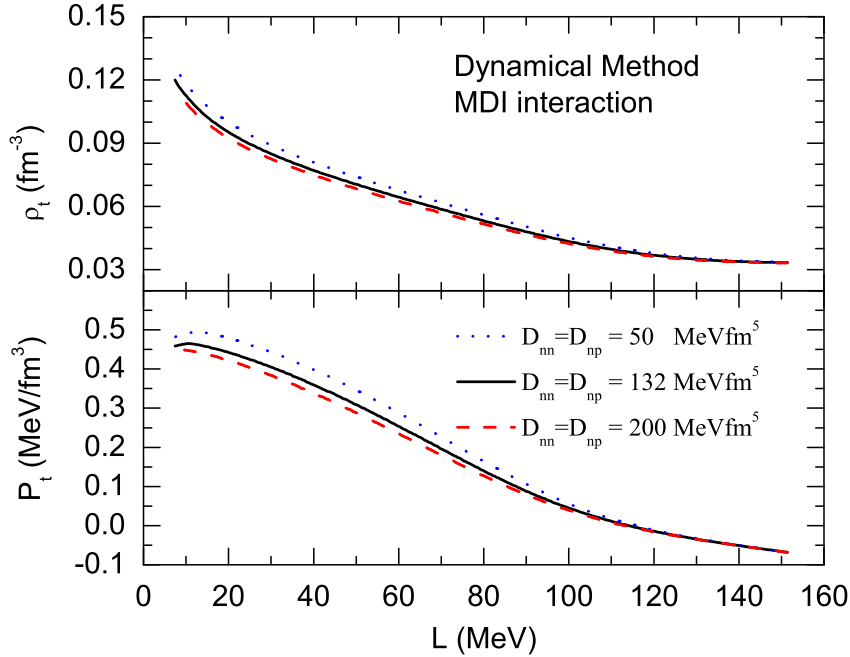}
\caption{(Color online) $\rho_t$ and $P_t$ as a function of $L$ by
using dynamical method for MDI interaction with different values of
$D_{pp}=D_{nn}=D_{np}=D_{pn}$.} \label{rhotPtD}
\end{figure}
In closing this subsection, we examine how the transition density
and pressure may be sensitive to variations of the coefficients
$D_{pp}=D_{nn}=D_{np}$ in the MDI interaction. As we have pointed
out in subsection~\ref{dynamical}, for the dynamical method, we
introduced phenomenologically the empirical values of
$D_{pp}=D_{nn}=D_{np}=132$ MeV$\cdot $fm$^{5}$ for the gradient
coefficients in the MDI interaction. These values are obviously not
obtained self-consistently. Shown in Fig.~\ref{rhotPtD} are the
$\rho_t$ and $P_{t}$ as functions of $L$ by using the dynamical
method with the full MDI EOS but different values of the
coefficients $D_{pp}=D_{nn}=D_{np}$, namely, $D_{pp}=D_{nn}=D_{np}=$
$50$, $132$, and $200$ MeV$\cdot $fm$^{5}$, respectively. We note
from Fig.~\ref{rhotPtD} that changing the value of
$D_{pp}=D_{nn}=D_{np}$ from $50$ to $200$ MeV$\cdot $fm$^{5}$ leads
to at most a variation of about 0.007 fm$^{-3}$ for $\rho_t$ and
0.06 MeV/fm$^{3}$ for $P_{t}$. These results thus indicate that the
transition density and pressure are rather insensitive to the
variation of $D_{pp}=D_{nn}$ and $D_{np}=D_{pn}$.

\subsection{Constructing the EOS from the center to the surface of neutron stars}
\label{ns-eos}

With a clear understanding about the core-crust transition density as we discussed above,
we now investigate several other properties of the crust and the whole neutron star.
To proceed, it is necessary to know the EOS over a broad density range from the center to the surface of
neutron stars. Besides the possible appearance of nuclear pasta in the crust,
various phase transitions and non-nucleonic degrees of freedom may
appear in the core. In this work, we restrict ourselves to the
simplest and traditional model. We make the minimum assumption that the core contains
the uniform $npe\mu$ matter only and there is no phase transition.
Results of this kind of calculations serve as a useful baseline for
understanding general features of astrophysical observations. Significant deviations from
observations may indicate the onset of non-traditional physics.

For the core we use the MDI EOS and its PA shown in Fig.\
\ref{MDInpemu}. In the inner crust of densities between $\rho_{out}$
and $\rho_t$ where the nuclear pastas may exist, because of our poor
knowledge about its EOS from first principle, following Carriere et
al.~\citep{Hor03} we construct its EOS according to
\begin{eqnarray}
P=a+b\epsilon^{4/3}. \label{crustEOS43}
\end{eqnarray}
This polytropic form with an index of $4/3$ has been found to be a
good approximation to the crust EOS~\citep{Lin99,Lat00} and we will
discuss how our results are sensitive to the polytropic index later.
The $\rho_{out}=2.46\times10^{-4}$ fm$^{-3}$ is the density
separating the inner from the outer crust. The constant $a$ and $b$
are determined by
\begin{eqnarray}
a &=& \frac{P_{out}\epsilon_t^{4/3}-P_t\epsilon_{out}^{4/3}}
{\epsilon_t^{4/3}-\epsilon_{out}^{4/3}}, \notag \\
b &=& \frac{P_t-P_{out}}{\epsilon_t^{4/3}-\epsilon_{out}^{4/3}},
\end{eqnarray}
where $P_t$, $\epsilon_t$ and $P_{out}$, $\epsilon_{out}$ are the
pressure and energy density at $\rho_t$ and $\rho_{out}$,
respectively. In the outer crust with $6.93\times10^{-13}$
fm$^{-3}<\rho<\rho_{out}$ we use the EOS of
BPS~\citep{BPS71,Iida97}, and in the region of $4.73\times10^{-15}$
fm$^{-3}<\rho<$$6.93\times10^{-13}$ fm$^{-3}$ we use the EOS of
FMT~\citep{BPS71}.

\begin{figure}[t!]
\centering
\includegraphics[scale=1.2]{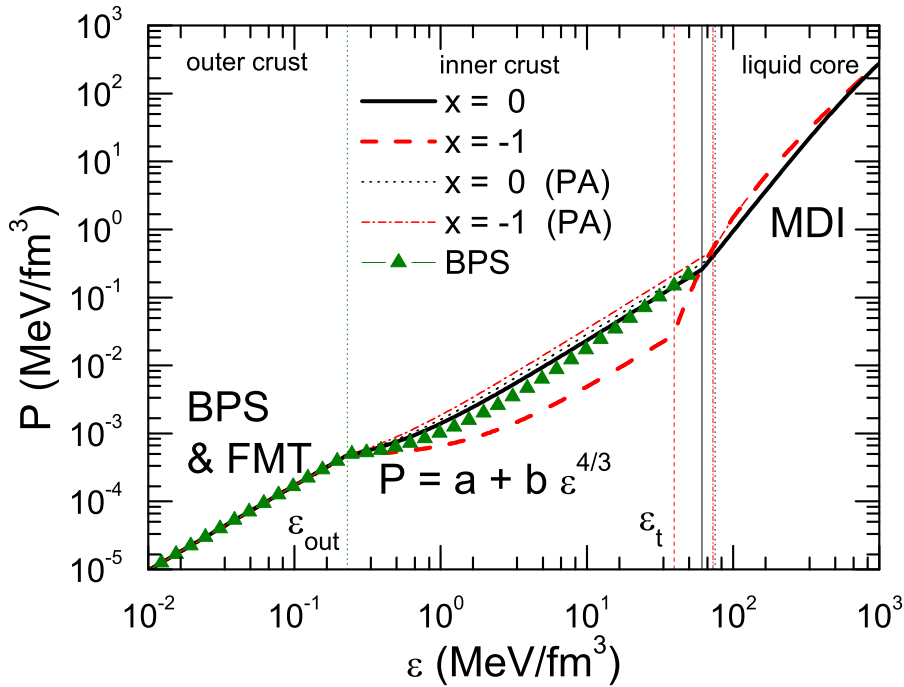}
\caption{{\protect\small (Color online) The EOS of different parts
of neutron stars. The energy density at $\rho_t$ and $\rho_{out}$ is
indicated in the figure as $\epsilon_t$ and $\epsilon_{out}$,
respectively, and the full and PA results of MDI interaction with
$x=0$ and $x=-1$ are shown.}} \label{Pepsilon}
\end{figure}

Shown in Fig.~\ref{Pepsilon} are the selected EOS for the different
parts of the neutron star. As we have discussed earlier, the
$\rho_t$ is obtained by studying the onset of instabilities in the
core, namely it is the critical density below which small density
fluctuations will grow exponentially. The $\rho_t$ is thus
determined by the EOS of the core only. We use here the $\rho_t$
obtained within the dynamical method using the full EOS and its
parabolic approximation with the MDI interaction of $x=0$ and
$x=-1$. The corresponding values of $\epsilon_t$ are indicated by
the vertical lines in Fig.~\ref{Pepsilon}. Using the above
combination of EOS's for the different parts of the neutron star,
the radial distribution of the total energy density and the pressure
in neutron stars is continuous as required, but the derivative of
the pressure is not continuous at $\rho_t$ and $\rho_{out}$. It is
seen that the EOS for the inner crust is quite different using the
Full EOS or its PA especially with $x=-1$. Interestingly, one can
see that the famous BPS EOS extended to the inner crust is between
the parameterized EOS's with $x=0$ and $x=-1$.

\subsection{The mass-radius correlation of neutron stars}
\label{nm-co}

With the EOS constructed above, in the next three subsections we
study several key properties of the crust and the whole neutron star
using the formalisms outline in section~\ref{nstarth}. We carry out
numerical calculations for all interested quantities. For the
crustal fraction of the moment of inertia, we also compare our
numerical calculations with predictions of the analytical expression
put forward by Lattimer and Prakash~\citep{Lat07,Lat00}. In this
subsection, we focus on effects of the symmetry energy on the
mass-radius correlation. We use the MDI interaction with $x=0$ and
$x=-1$ consistent with the existing heavy-ion reaction
data~\citep{LCK08}.

\begin{figure}[t!]
\centering
\includegraphics[scale=1.2]{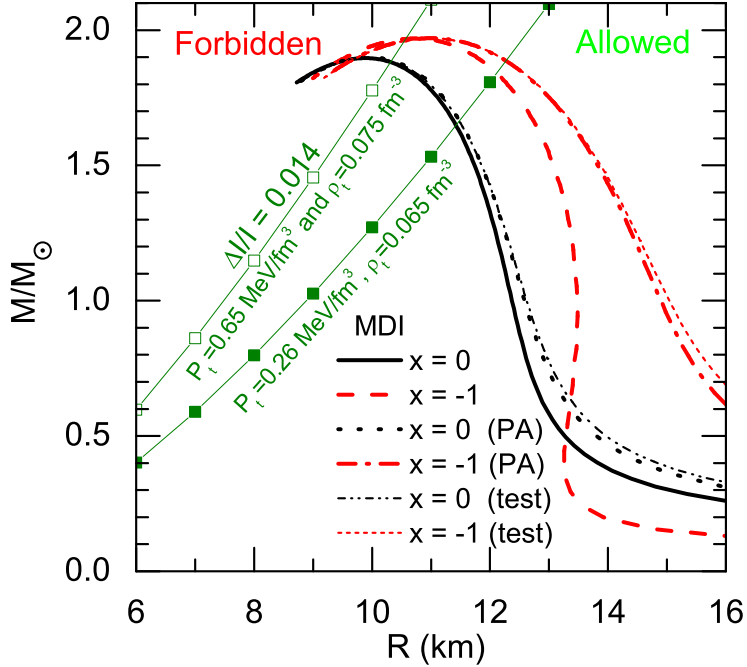}
\caption{(Color online) The $M$-$R$ relation of static neutron stars
from the full EOS and its parabolic approximation as well as the
test case (see text for details) with the MDI interaction with $x=0$
and $x=-1$. For the Vela pulsar, the constraint of $\Delta
I/I>0.014$ implies that allowed masses and radii lie to the right of
the line linked with solid squares ($\protect\rho _{t}=0.065$
fm$^{-3}$ and $P_{t}=0.26$ MeV/fm$^{3}$, obtained in the present
work) or open squares ($\protect\rho _{t}=0.075$ fm$^{-3}$ and
$P_{t}=0.65$ MeV/fm$^{3}$, used in ref.~\protect\citep{Lin99}).}
\label{MR}
\end{figure}

\begin{figure}[t!]
\centering
\includegraphics[scale=1.2]{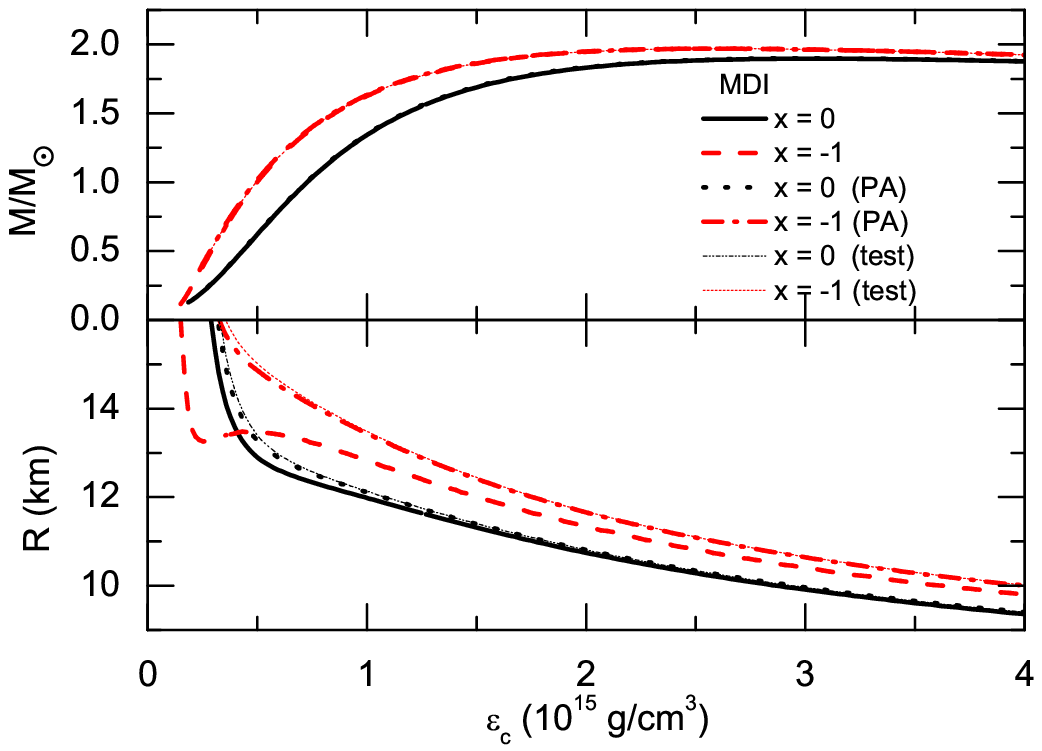}
\caption{{\protect\small (Color online) The mass and radius of neutron stars as
functions of the central energy density using the
MDI interaction with $x=0$ and $x=-1$. The results from three
methods are shown for comparison. See text for details. }}
\label{epsiloncMR}
\end{figure}

The resulting mass-radius correlation is shown in Fig.~\ref{MR}. For
the softer symmetry energy ($x=0$) the $M$ decreases with increasing
$R$, while for the stiffer symmetry energy($x=-1$) the radius
remains almost unchanged or even decreases with decreasing mass near
$R=13.5$ km. For $M>0.53 M_{\odot}$ the radius is larger for $x=-1$,
while for $M<0.53 M_{\odot}$ the radius is larger for $x=0$. For
nucleonic matter, a stiffer symmetry energy leads to a stiffer EOS
for the liquid core, but a lower core-crust transition density. The
crossing point of the M-R curves with $x=0$ and $x=-1$ is a result
of this competition. It is clearly shown that with the PA the radius
is larger at a fixed mass especially for the stiffer symmetry energy
of $x=-1$. To better understand the role of the transition density
in determining the M-R relation, we also made an additional test by
using the full MDI EOS but with the $\rho_t$ obtained from using the
PA. The results are shown with the dotted lines. They are very close
to the results obtained consistently using the PA in calculating
both the EOS and the transition density. Thus, the mass-radius
relation, especially the radius, seems to be quite sensitive to the
location of the inner edge. The small difference between the full
EOS and its PA for the core (shown in the panels (a) and (b) of
Fig.~\ref{MDInpemu}) has a negligible effect on the M-R relationship
once the inner edge is fixed. These features can be seen more
clearly in Fig.~\ref{epsiloncMR} where the mass and radius are
displayed separately as functions of the central energy density.
Very similar masses are obtained independently of how the $\rho_t$
was calculated for a given $x$ parameter. However, the radii are
appreciably different for the stiffer symmetry energy with $x=-1$
using the full EOS or its PA because of their different $\rho_t$
values. Also, since the test case has the same $\rho_t$ as the PA,
it thus leads to the same radii as the PA for both $x=0$ and $x=-1$.

\subsection{The crust thickness and the core size of neutron stars}
\label{dr-ns}

\begin{figure}[t!]
\centering
\includegraphics[scale=1.2]{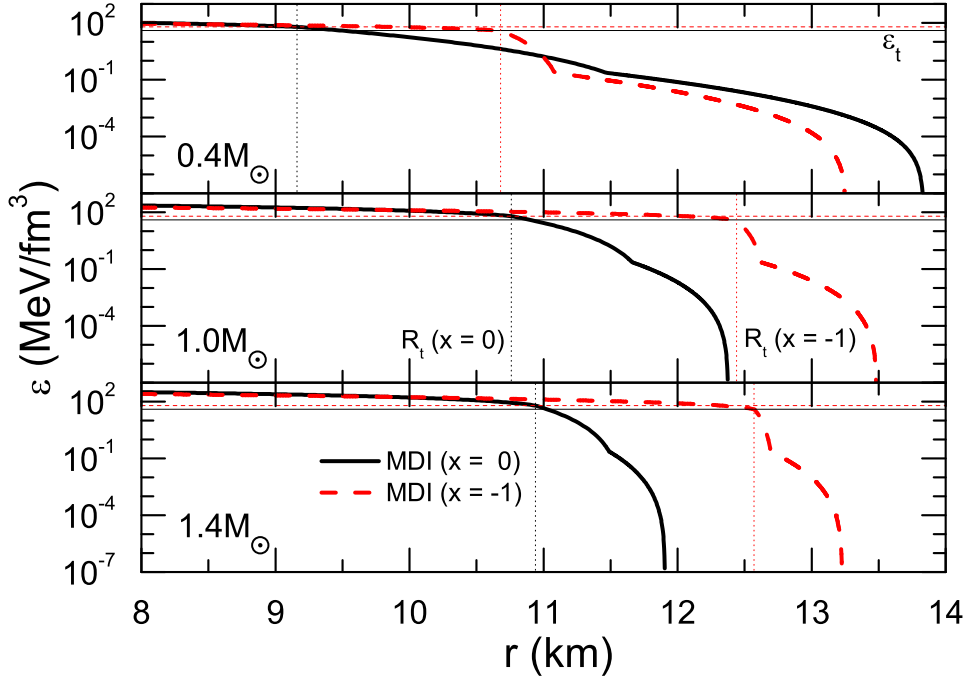}
\caption{{\protect\small (Color online) The radial energy density
distribution of neutron stars, using MDI interaction with $x=0$ and
$x=-1$, at total mass $0.4M_{\odot}$, $1.0M_{\odot}$ and
$1.4M_{\odot}$. The interface between uniform part and crust part is
indicated. $R_t$ is the radius of the liquid core and $\epsilon_t$
is the energy density at the edge of the liquid core and the
crust.}} \label{epsilonr}
\end{figure}

For a given neutron star of total mass $M$ and radius $R$, what are
the respective sizes of its core and crust? How do they depend on
the stiffness of the symmetry energy? How do they depend on the
neutron star mass $M$? It is well known that the size of neutron
skin in heavy nuclei increases with the increasing $L$ as shown in
Fig.\ \ref{SrhotL}. How does the thickness of neutron star crusts
depend on the $L$? These are among the interesting questions we
shall discuss in this subsection. First, we display in
Fig.~\ref{epsilonr} the radial energy density profile for neutron
stars of total mass $0.4M_{\odot}$, $1.0M_{\odot}$ and
$1.4M_{\odot}$ using the MDI interaction with $x=0$ and $x=-1$,
respectively. The inner edge separating the uniform core from the
crust is indicated by the vertically dotted lines. The corresponding
energy density $\epsilon_t$ is shown as the longitudinally dotted
lines. It is very interesting to see that the radius of the core
increases while the thickness of the crust decreases with the
increasing neutron star mass $M$. The lighter neutron stars
generally have thicker and more diffusive crusts due to the
competition between the gravitation and the nuclear forces.
Moreover, this feature is independent of the symmetry energy used.
It is also seen that the stiffer symmetry energy with $x=-1$
predicts a larger core but a thinner crust for a given mass $M$.
More quantitatively, for a canonical neutron star of
$M=1.4M_{\odot}$, the radius of the core is $10.89$ km with $x=0$
and $12.55$ km with $x=-1$, and the thickness of the crust is $1.09$
km with $x=0$ and $0.72$ km with $x=-1$, respectively. Therefore,
with a softer symmetry energy, a light neutron star can have a big
radius due to its very thick crust.

\begin{figure}[t!]
\centering
\includegraphics[scale=1.2]{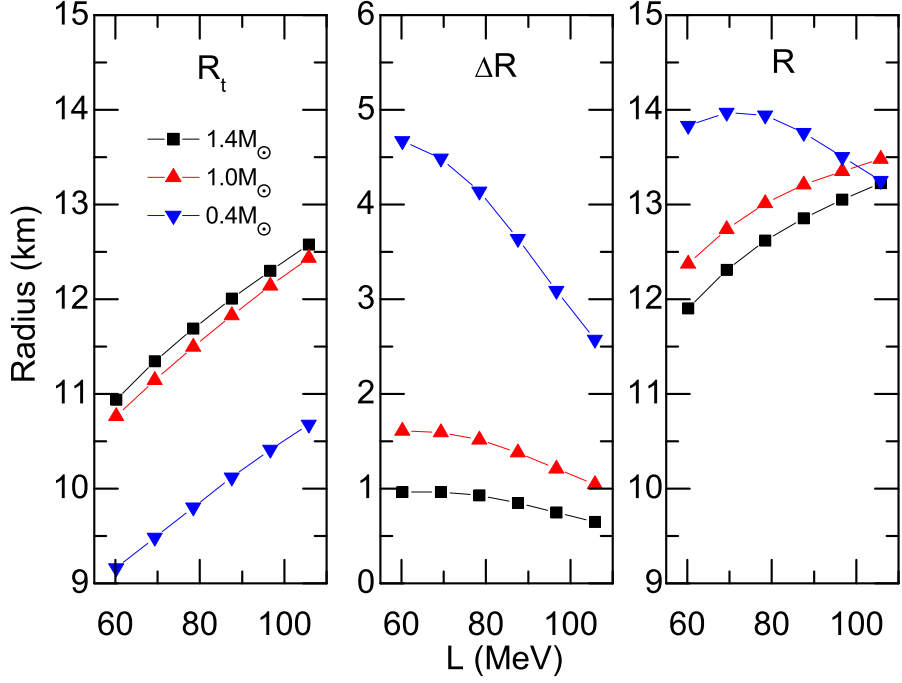}
\caption{{\protect\small (Color online) The whole radius $R$, the crust thickness
$\Delta R$, the core radius $R_t$ as functions of $L$ at fixed total mass of $0.4M_{\odot}$,
$1.0M_{\odot}$ and $1.4M_{\odot}$, respectively.}}\label{RL}
\end{figure}

To study more systematically effects of the symmetry energy, we show
in Fig.~\ref{RL} the core radius $R_t$, the crust thickness $\Delta
R$ and the total radius $R$ as functions of $L$ for a fixed total
mass of $0.4M_{\odot}$, $1.0M_{\odot}$ and $1.4M_{\odot}$,
respectively. It is seen that the $R_t$ increases almost linearly
with the increasing $L$. The $R_t$ also increases with the
increasing mass at a fixed $L$. This is because the stiffer the
symmetry energy is, the larger the contribution of the isospin
asymmetric part of the pressure will be, which makes the $R_t$
larger. Moreover, the $\Delta R$ decreases with the increasing $L$
especially for light neutron stars, as the transition density
decreases with the increasing $L$. As the thickness of the crust
$\Delta R$ and the core radius $R_t$ depend oppositely on $L$, the
total radius $R=R_t+\Delta R$ of the neutron star may show a
complicated dependence on $L$. We stress here that this is the
result of a competition between the repulsive nuclear pressure
dominated by the symmetry energy contribution and the gravitational
binding. Interestingly, it is often mentioned that the crust of
neutron stars bears some analogy with the neutron-skin of heavy
nuclei. However, they show completely opposite dependences on the
$L$. Namely, the size of neutron-skin usually increases with the
increasing $L$ as a result of the competition between the nuclear
surface tension and the pressure difference of neutrons and protons,
while the thickness of neutron star crusts decreases with the
increasing $L$ as a result of the competition between the nuclear
pressure and the gravitational binding.

\subsection{The crustal fractions of neutron star masses and moments of inertia}
\label{di-ns}

\begin{figure}[t!]
\centering
\includegraphics[scale=1.2]{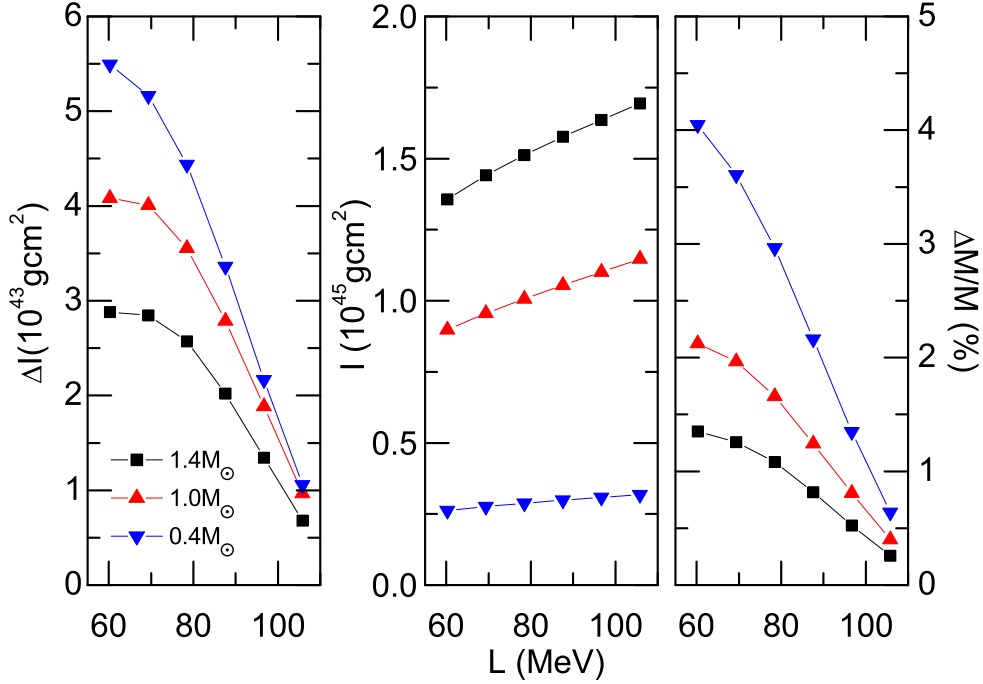}
\caption{{\protect\small (Color online) The crustal fraction of
neutron mass $\Delta M/M$, the moment of inertia $I$ of the whole
star and the crust contribution $\Delta I$ as a function of $L$, at
fixed total mass $0.4M_{\odot}$, $1.0M_{\odot}$ and $1.4M_{\odot}$,
respectively.}}\label{MIL}
\end{figure}

What is the crustal fraction $\Delta M/M$ of the total mass and how
does it depend on the symmetry energy? Since the mass is simply the
integration of the energy density, from the profile of the energy
density shown in Fig.\ \ref{epsilonr} we expect the $\Delta M/M$ and
$\Delta R/R$ have very similar dependences on $L$. Shown in the
right window of Fig.~\ref{MIL} is the $\Delta M/M$. The fractional
mass of the crust decreases with the increasing $L$ at a fixed total
mass, and it decreases with the increasing total mass at a fixed
value of $L$. The moment of inertia is determined by the
distribution of the energy density. From the middle window, it is
seen that the total moment of inertia increases with the increasing
mass at a fixed value of $L$ and increases with the increasing $L$
at a fixed total mass. The dependence on $L$ is relatively weak
especially for the light neutron stars. However, the crust
contribution of the moment of inertia varies much more quickly with
$L$. It decreases with the increasing neutron star mass at a fixed
value of $L$ and decreases with the increasing $L$ at a fixed total
mass. These are all consistent with the behaviors of the fractional
mass and size of the crust.

\begin{figure}[t!]
\centering
\includegraphics[scale=1.2]{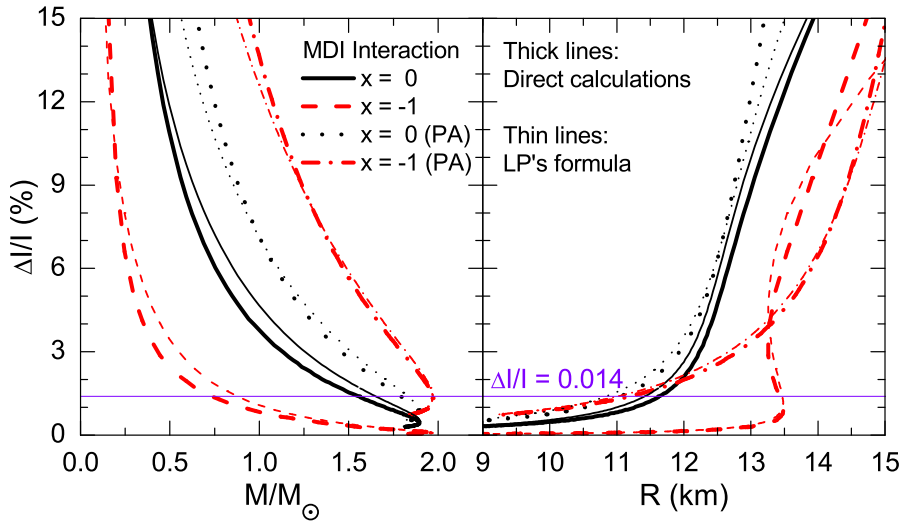}
\caption{{\protect\small (Color online) The relation between the
crustal fraction of the moment of inertia and the total mass or the
radius of neutron stars, using MDI interaction with $x=0$ and
$x=-1$. The results of PA or exact method from direct calculation or
LP's formula are shown for comparison, and the constraint of $\Delta
I/I$ is also indicated. }} \label{dIMR}
\end{figure}

The crustal fraction of the moment of inertia ${\Delta I}/{I}$ is
particularly interesting as it can be inferred from observations of
pulsar glitches, the occasional disruptions of the otherwise
extremely regular pulsations from magnetized, rotating neutron stars.
It can be expressed approximately as~\citep{Lat07,Lat00}
\begin{eqnarray}\label{dI}
\frac{\Delta I}{I} &=& \frac{28\pi P_t R^3}{3 M c^2}
\frac{(1-1.67\xi-0.6\xi^2)}{\xi}
\left[1+\frac{2P_t(1+5\xi-14\xi^2)}{\rho_t m c^2 \xi^2}\right]^{-1},
\end{eqnarray}
where $m$ is the mass of baryons and $\xi=G M/R c^2$. A numerical
verification of this formula is useful. Predictions of this formula
(thin lines) are compared in Fig.~\ref{dIMR} with our direct
numerical calculations (thick lines). Very interestingly, the
analytical formula reproduces very well our results from direct
numerical calculations using both the full EOS and its PA. Comparing
calculations using the full EOS and its PA, one sees clearly big
differences, again due to the corresponding differences in the
transition density. For instance, using either the direct numerical
calculation or the formula (\ref{dI}), at a fixed total mass $M$ the
$\Delta I/I$ increases using the full EOS while it decreases using
the PA when the $x$ parameter is changed from $x=-1$ to $x=0$. As it
was stressed in ref.~\citep{Lat00}, the $\Delta I/I$ depends
sensitively on the symmetry energy at sub-saturation densities
through the $P_t$ and $\rho_t$, but there is no explicit dependence
upon the EOS at higher-densities.

Experimentally, the crustal fraction of the moment of inertia has
been constrained as ${\Delta I}/{I}>0.014$ from studying the
glitches of the Vela pulsar~\citep{Lin99}. As indicated in
Fig.~\ref{dIMR}, this limits the masses and radii of the neutron
star. For example, from Fig.~\ref{dIMR}, it is indicated that the
maximum mass is about $1.57M_{\odot }$ ($0.73M_{\odot }$) while its
minimum radius is about $11.6$ ($13.4$) km for the MDI interaction
with $x=0$ ($x=-1$) if the dynamical method is used to determine the
$P_t$ and $\rho_t$. We note here that the very small mass for Vela
pulsar constrained by this condition using the MDI with $x=-1$ is
due to the associated small transition density and pressure.
Combining the observational constraint of $\Delta I/I>0.014$ with
the upper bounds of $\rho _{t}=0.065$ fm$^{-3}$ and $P_{t}=0.26$
MeV/fm$^{3}$ inferred from heavy-ion reactions, we can obtain a
minimum radius of $R\geq 4.7+4.0M/M_{\odot }$ km for the Vela
pulsar. This limit is indicated by the solid squares in
Fig.~\ref{MR}. According to this constraint, the radius of the Vela
pulsar is predicted to exceed $10.5$ km should it have a mass of
$1.4M_{\odot }$. It is worth mentioning that a constraint of $R\geq
3.6+3.9M/M_{\odot }$ km for this pulsar (see the open squares in
Fig.~\ref{MR}) has been derived previously in ref.~\citep{Lin99} by
using $\rho _{t}=0.075$ fm$^{-3}$ and $P_{t}=0.65$ MeV/fm$^{3}$. The
difference between this and our prediction is due to the different
$\rho_t$ and $P_{t}$.

\subsection{The inner crust EOS dependence of neutron star properties}
\label{crustEOS}

As discussed in Eq. (\ref{crustEOS43}) of subsection~\ref{ns-eos}, we
have adopted the polytropic EOS of $P=a+b\epsilon^{\gamma }$ with $\gamma = 4/3$
for the inner crust in the above calculations. This particular polytropic EOS
has been extensively used for studying the inner crust in
the literature~\citep{Lin99,Lat00,Hor03}. However, due to the complexity of
the inner crust, its EOS is rather uncertain~\citep{Neg73}. Thus, it would be
interesting to investigate how our results may be sensitive to the polytropic
index $\gamma$.

\begin{figure}[t!]
\centering
\includegraphics[scale=1.2]{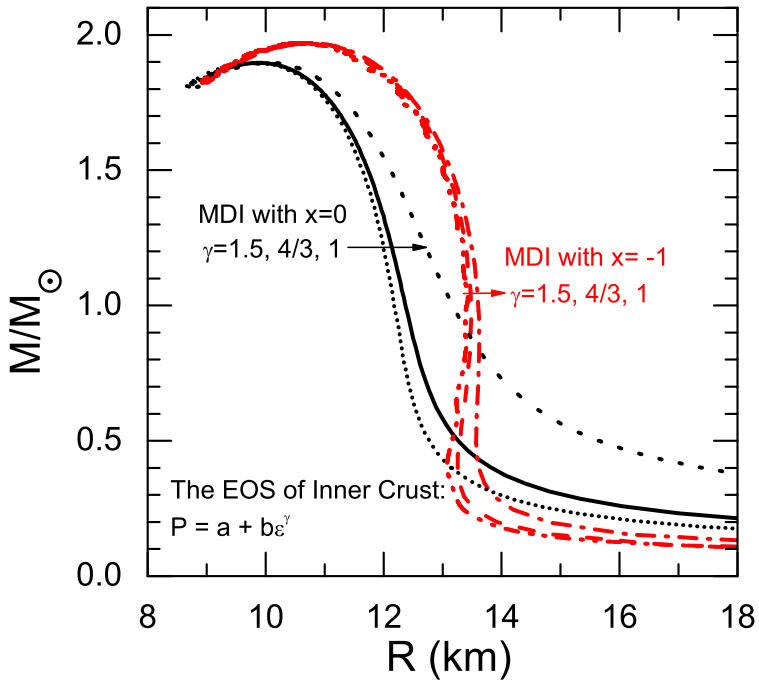}
\caption{(Color online) The $M$-$R$ relation of static neutron stars
from the full MDI EOS with $x=0$ and $x=-1$. The different values of
the polytropic index $\gamma $ in $P=a+b\epsilon^{\gamma }$ for the
inner crust EOS, i.e., $\gamma =1.5, 4/3$, and $1$ have been used.}
\label{MRcrustEOS}
\end{figure}
Firstly, let us see how the polytropic index affects the mass and
radius of a neutron star. Shown in Fig.~\ref{MRcrustEOS} is the
$M$-$R$ relation obtained using the full MDI EOS with $x=0$ and
$x=-1$ with $3$ different values of the polytropic index $\gamma $,
i.e., $\gamma =1.5, 4/3$, and $1$. It is seen that the polytropic
index has very little effects on the mass of neutron stars. On the
other hand, it is interesting to see that the neutron star radius
increases significantly with the deceasing polytropic index $\gamma
$ especially for the softer symmetry energy ($x=0$). This is due to
the change of the crust thickness from varying the inner crust EOS.
The observed symmetry energy dependence of the polytropic index
effects on the neutron star radius can be easily understood since
the stiffer symmetry energy leads to a thinner thickness of the
crust as shown in Fig.~\ref{RL} and thus less sensitivity to the
variation of the inner crust EOS. Our results thus indicate that for
softer symmetry energies, an accurate inner crust EOS is important
for the precise determination of the neutron star radius.

\begin{figure}[t!]
\centering
\includegraphics[scale=1.2]{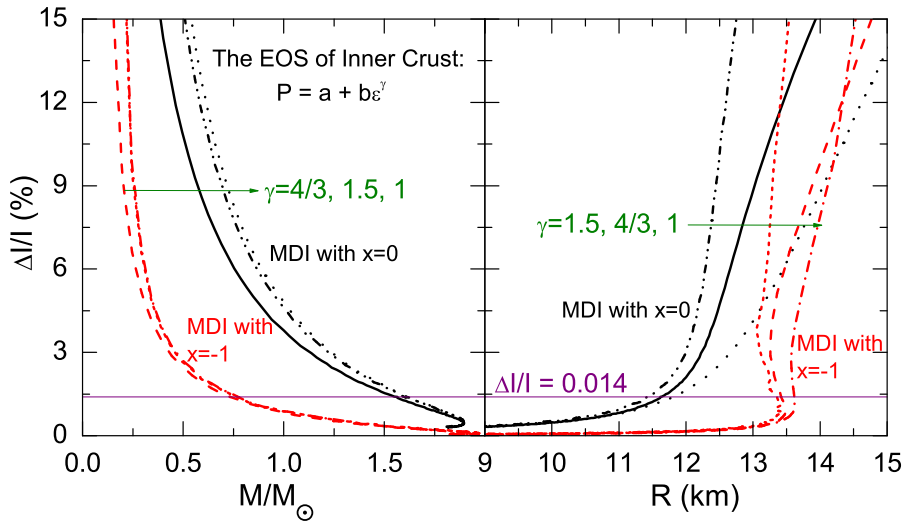}
\caption{{\protect\small (Color online) The relation between the
crustal fraction of the moment of inertia and the total mass or the
radius of neutron stars, using MDI interaction with $x=0$ and
$x=-1$. The different values of the polytropic index $\gamma $ in
$P=a+b\epsilon^{\gamma }$ for the inner crust EOS, i.e., $\gamma
=1.5, 4/3$, and $1$ have been used. The constraint of $\Delta I/I$
is also indicated.}} \label{dIMRcrustEOS}
\end{figure}
In order to see the inner crust EOS dependence of the crustal
fraction of the moment of inertia ${\Delta I}/{I}$, we show in
Fig.~\ref{dIMRcrustEOS} the ${\Delta I}/{I}$ as functions of the
total mass and the radius of neutron stars from the full MDI EOS
with $x=0$ and $x=-1$ using $\gamma =1.5, 4/3$, and $1$. Noting the
very weak dependence of the neutron star mass on the $\gamma $
index, we can see clearly from the left window of
Fig.~\ref{dIMRcrustEOS} that the ${\Delta I}/{I}$ is not so
sensitive to the variation of the inner crust EOS, especially for
stiffer symmetry energies. We notice here that the pronounced
$\gamma $ index dependence of ${\Delta I}/{I}$ as a function of the
neutron star radius is due to the fact that the neutron star radius
depends significantly on the $\gamma $ index as shown in
Fig.~\ref{MRcrustEOS}, especially for the softer symmetry energy
($x=0$).

Also indicated in Fig.~\ref{dIMRcrustEOS} is the constraint of
$\Delta I/I$ from studying the glitches of the Vela
pulsar~\citep{Lin99}. It is very interesting to see that the neutron
star mass and radius at $\Delta I/I=0.014$ exhibit a very weak
dependence on the polytropic index $\gamma $. This nice feature
implies that the obtained constraint on the minimum radius of $R\geq
4.7+4.0M/M_{\odot }$ km for the Vela pulsar in the present work is
not sensitive to the inner crust EOS and still holds. Moreover, we
can see from Fig.~\ref{dIMRcrustEOS} that the robustness of the
constraint $R\geq 4.7+4.0M/M_{\odot }$ km against the variation of
the inner crust EOS is actually due to the very small value of
$\Delta I/I$ for the Vela pulsar. For higher values of $\Delta I/I$,
on the contrary, the constraint will depend significantly on the
inner crust EOS.

\section{Summary}
\label{summary}

In summary, we first analyzed the relationship between the well
established dynamical and thermodynamical methods for locating the
inner edge separating the uniform liquid core from the solid crust
in neutron stars. It is shown analytically that the thermodynamical
method corresponds to the long-wavelength limit of the dynamical one
when the Coulomb interaction is neglected. Moreover, it is shown
that the results obtained from using the full expression of the EOS
for a given interaction are very similar for the two methods.
However, the widely used parabolic approximation to the EOS of
asymmetric nuclear matter leads systematically to significantly
higher core-crust transition densities and pressures, especially for
stiffer symmetry energy functionals regardless of the specific
method used in calculating the transition density. Our results thus
indicate that one can hardly obtain the accurate transition density
without knowing the complete EOS and may introduce a huge error by
assuming {\it a priori} that the EOS is parabolic in isospin
asymmetry for a given interaction. Based on systematical
calculations using the modified Gogny force (MDI interaction) and
selected $51$ Skyrme interactions widely used in the literature, it
is shown that the transition density and pressure are very sensitive
to the density dependence of the nuclear symmetry energy. We also
systematically investigated several properties of neutron star
crust. We found that the thickness, fractional mass and moment of
inertia of neutron star crust are all very sensitive to the slope
parameter $L$ of the nuclear symmetry energy through the transition
density $\rho _{t}$ and the results depend on whether one uses the
full EOS or its parabolic approximation. Therefore, accurate
knowledge on the nuclear symmetry energy at sub-saturation densities
is required to fully understand the properties of neutron star
crusts.

Using the MDI EOS of neutron-rich nuclear matter constrained by the
recent isospin diffusion data from heavy-ion reactions in the same
sub-saturation density range as the neutron star crust, the
transition density and pressure at the inner edge of neutron star
crusts are limited to $0.040$ fm$^{-3}$ $\leq \rho _{t}\leq 0.065$
fm$^{-3}$ and $0.01$ MeV/fm$^{3}$ $\leq P_{t}\leq 0.26$
MeV/fm$^{3}$, respectively. The constrained range of the transition
density is significantly below the fiducial value of $\rho_t\approx
0.08$ fm$^{-3}$ often used in the literature and the estimate of
$0.5<\rho_t/\rho_0<0.7$ made previously within the thermodynamical
approach using the parabolic approximation of the EOS while that of
the $P_{t}$ is also significantly less than the fiducial value of
$P_t\approx 0.65$ MeV/fm$^{3}$ often used in the literature. The
newly constrained transition density and pressure together with the
condition $\Delta I/I>0.014$ for the crustal fraction of the moment
of inertia extracted from studying glitches of the Vela pulsar allow
us to set a new limit on the radius of the Vela pulsar, i.e., $R\geq
4.7+4.0M/M_{\odot }$. It is significantly different from the
previous estimate and thus puts a new constraint for the mass-radius
relation of neutron stars.

Finally, it is worth noting that in the present work, we have only
considered the non-accreting crusts of cold, non-rotating nucleonic
neutron stars. In the next step, we plan to extend the study to
accreting neutron stars. It will be especially interesting to
investigate how the finite temperature, the strong magnetic field
and neutrino trapping may affect the transition density and pressure
reported here. Moreover, there are still many interesting issues
regarding the neutron star crust, such as its composition, thermal,
transport and mechanical properties that are important for a better
understanding of the structure and evolution of protoneutron stars,
the x-ray bursts and the emission of gravitational waves from
neutron stars. More information from terrestrial nuclear reactions
especially those induced by radioactive beams will certainly
contribute to resolving these issues.

\section*{Acknowledgements}This work was supported in part by the National Natural Science
Foundation of China under Grant Nos. 10575071, 10675082, and
10874111, MOE of China under project NCET-05-0392, Shanghai
Rising-Star Program under Grant No. 06QA14024, the SRF for ROCS, SEM
of China, the National Basic Research Program of China (973 Program)
under Contract No. 2007CB815004, the US National Science Foundation
under Grant No. PHY-0652548, PHY-0757839 and PHY-0457265, the Welch
Foundation under Grant No. A-1358, the Research Corporation under
Award No. 7123 and the Texas Coordinating Board of Higher Education
Award No. 003565-0004-2007.


\begin{thebibliography}{}

\bibitem[{Arpoen (1972)}] {Arp72} Arpoen, J. 1972, Nucl. Phys. A, 191, 257

\bibitem[{Avancini {et~al.} (2008a)}] {Ava08} Avancini, S.S., et al. 2008a, Phys. Rev. C, 78, 015802

\bibitem[{Avancini {et~al.} (2008b)}] {Ava08b} Avancini, S.S., et al. 2008b, arXiv:0812.3170v1 [nucl-th]

\bibitem[{Baym {et~al.} (1971a)}] {BPS71} Baym, G., Pethick, C. J., \& Sutherland, P. 1971a, ApJ, 170, 299

\bibitem[{Baym {et~al.} (1971b)}] {BBP71} Baym, G., Bethe, H. A., \& Pethick, C. J. 1971b, Nucl.
Phys. A, 175, 225

\bibitem[{Bombaci \& Lombardo (1991)}] {Bom91} Bombaci, I. \& Lombardo, U., 1991, Phys. Rev. C, 44, 1892

\bibitem[{Bombaci (2001)}]{Bom01}Bombaci, I., in Isospin Physics in Heavy-Ion Collisions at
Intermediate Energies, eds. Bao-An Li and W. Udo Schroder (Nova
Science Publishers, Inc., New York, 2001), p.35

\bibitem[{Brack {et~al.} (1985)}] {Bra85} Brack, M., Guet, C. \& Hakansson H.-B. 1985, Phys. Rept., 123,
275

\bibitem[{Brown (1998)}] {Bro98} Brown, B.A. 1998, Phys. Rev. C, 58, 220.

\bibitem[{Brown (2000)}] {Brown00} Brown, B.A. 2000, Phys. Rev. Lett., 85,
5296

\bibitem[{Burrows {et~al.} (2006)}] {Bur06} Burrows, A., Reddy, S., \& Thompson, T. A. 2006, Nucl. Phys.
A, 777, 356

\bibitem[{Carriere {et~al.} (2003)}] {Hor03} Carriere, J., Horowitz, C.J., \& Piekarewicz,
J. 2003, ApJ, 593, 463

\bibitem[{Chabanat {et~al.} (1997)}] {Chabanat} Chabanat, E., Bonche, E., Haensel,
E., Meyer, J., \& Schaeffer, R. 1997, Nucl. Phys. A, 627, 710

\bibitem[{Chamel {et~al.} (2008a)}] {Cha08a} Chamel, N., Goriely, S. \& Pearson, J.M.
2008, Nucl. Phys. A, 812, 72

\bibitem[{Chamel \& Haensel (2008)}] {Cha08} Chamel, N \& Haensel, P. 2008, Living Rev. Relativity, 11, 10

\bibitem[{Chen {et~al.} (2001)}] {Che01} Chen, L.~W., Zhang, F.~S., Lu, Z.~H., Li, W.~F., Zhu,
Z.~Y., \& Ma, H.~R. 2001, J. Phys. G, 27, 1799

\bibitem[{Chen {et~al.} (2005a)}] {Che05a} Chen, L.~W., Ko, C.~M., \& Li, B.~A. 2005a, Phys. Rev. Lett., 94,
032701

\bibitem[{Chen {et~al.} (2005b)}] {Che05b} Chen, L.~W., Ko, C.~M., \& Li, B.~A. 2005b, Phys. Rev. C, 72,
064309

\bibitem[{Chen {et~al.} (2007)}] {Chen07} Chen, L.~W., Ko, C.~M., \& Li, B.~A.  2007, Phys. Rev. C, 76,
054316

\bibitem[{Chomaz {et~ al.}(2004)}]{Cho04} Chomaz, Ph., Colonna, M., Randrup, J., 2004, Phys.
Rep., 389, 263

\bibitem[{Das {et~al.} (2003)}] {Das03} Das, C.B., Gupta, S.~D., Gale, C., \& Li, B.-A. 2003, Phys. Rev.,
C, 67, 034611

\bibitem[{Das {et~al.}(2005)}]{Das05} Das, C.B., Das Gupta, S., Lynch, W.G.,
Mekjian,A.Z.,  Tsang, M.B., 2005, Phys. Rep., 406, 1

\bibitem[{Dieperink {et~al.}(2003)}]{Die03} Dieperink, A.E.L., et al. 2003, Phys. Rev. C, 68, 064307

\bibitem[{Douchin \& Haensel (2000)}] {Dou00} Douchin, F. \& Haensel, P. 2000, Phys. Lett.
B, 485, 107

\bibitem[{Douchin \& Haensel (2001)}] {Dou01} Douchin, F., \& Haensel, P. 2001, A\&A, 380, 151

\bibitem[{Ducoin {et~al.} (2007)}] {Chomaz07} Ducoin, C., Chomaz, Ph., \& Gulminelli, F. 2007, Nucl.
Phys. A, 789, 403

\bibitem[{Duncan (1998)}] {Dun98} Duncan, R. C. 1998, ApJL, 498, L45

\bibitem[{Friedrich \& Reinhard (1986)}] {Fri86} Friedrich J. \& Reinhard P.-G. 1986, Phys. Rev. C, 33, 335

\bibitem[{Furnstahl (2002)}]{Fur02}Furnstahl, R.J. 2002, Nucl. Phys. A, 706, 85

\bibitem[{Gale {et~ al.} (1987)}]{GBD87} Gale, C., Bertsch, G.F., \& Das Gupta, S., 1987, Phys. Rev. C, 35,
1666

\bibitem[{Gale {et~ al.} (1990)}]{Gal90} Gale, C., Welke, G. M., Prakash, M., Lee, S. J., \& Das
Gupta, S., 1990, Phys. Rev. C, 41, 1545

\bibitem[{Gogelein {et~al.} (2008)}] {Gog08} G\"{o}gelein, P., van Dalen, E. N. E.,
Fuchs, C., \& M\"{u}ther, H. 2008, Phys. Rev. C, 77, 025802

\bibitem[{Goriely {et~al.} (2003)}] {Gor03} Goriely, S., et al.
2003, Phys. Rev. C, 68, 054325

\bibitem[{Goriely {et~al.} (2005)}] {Gor05} Goriely, S., et al.
2005, Nucl. Phys. A, 750, 425

\bibitem[{Goriely {et~al.} (2006)}] {Gor06} Goriely, S., Samyn, M.
\& Pearson, J.M. 2006, Nucl. Phys. A, 773, 279

\bibitem[{Goriely {et~al.} (2007)}] {Gor07} Goriely, S., Samyn, M.
\& Pearson, J.M. 2007, Phys. Rev. C, 75, 064312

\bibitem[{Goriely \& Pearson (2008)}] {Gor08} Goriely, S. \& Pearson, J.M.
2008, Phys. Rev. C, 77, 031301(R)

\bibitem[{Hashimoto {et~al.} (1984)}] {Has84} Hashimoto, M., Seki, H., \& Yamada, M. 1984, Prog. Theor.
Phys., 71, 320

\bibitem[{Hempel \& Schaffner-Bielich (2008)}] {Hem08} Hempel, M. \& Schaffner-Bielich, J. 2008, J. Phys.
G, 35, 014043

\bibitem[{Horowitz \& Piekarewicz (2001)(2002)}] {Hor01} Horowitz, C.J. \& Piekarewicz, J. 2001,
Phys. Rev. Lett., 86 , 5647; 2001, Phys. Rev. C, 64, 062802 (R);
2002, Phys. Rev. C, 66, 055803

\bibitem[{Horowitz {et~al.}(2001)Horowitz, Pollock, Souder, Michaels {et~al.}}]{Horowitz:2001}
Horowitz, C.~J., Pollock, S. J., Souder, P. A., \& Michaels, R.
2001, Phys. Rev. C, 63, 025501

\bibitem[{Horowitz {et~al.} (2004)(2004)}] {Hor04} Horowitz, C. J., et al. 2004, Phys. Rev. C, 69, 045804;
Horowitz, C. J., et al. 2004, Phys. Rev. C, 70, 065806

\bibitem[{Horowitz (2005)}]{Hor05} Horowitz, C.J., talk at the World Consensus
Initiative, 12-16 February 2005, College Station, Texas, USA.
http://cyclotron.tamu.edu/wci3/

\bibitem[{Horowitz (2006)}]{Hor06} Horowitz, C.J. 2006, Eur. Phys. J. A, 30, 303

\bibitem[{Iida \& Sato (1997)}] {Iida97} Iida, K., Sato, K. 1997, ApJ, 477, 294

\bibitem[{Krastev \& Li (2007)}]{Kra07}Krastev, P.G. \& Li, B.A., 2007, Phys. Rev. C, 76, 055804.

\bibitem[{Krastev {et al.} (2008a)}]{Kra08a}Krastev, P.G., Li, B.A., \& Worley, A., 2008a,
ApJ, 676, 1170

\bibitem[{Krastev {et al.} (2008b)}]{Kra08b}Krastev, P.G., Li, B.A., \& Worley, A., 2008b,
Phys. Lett. B, 668, 1

\bibitem[{Kubis (2007)(2004)}] {Kub07} Kubis, S. 2007, Phys. Rev. C, 76
, 035801; 2004, Phys. Rev. C, 70, 065804

\bibitem[{Lattimer \& Prakash (2000)(2001)}] {Lat00} Lattimer, J.M. \& Prakash, M. 2007, Phys. Rep., 333,
121; 2001, ApJ, 550, 426

\bibitem[{Lattimer \& Prakash (2004)}] {Lat04} Lattimer, J.M. \& Prakash, M. 2004, Science, 304, 536

\bibitem[{Lattimer \& Prakash (2007)}] {Lat07} Lattimer, J.M. \& Prakash, M. 2007, Phys. Rep., 442, 109

\bibitem[{Li \& Ko (1997)}]{LiKo97}Li, B.A. \& Ko, C.M. 1997, Nucl. Phys A, 618,
498

\bibitem[{Li \& Chen (2005)}] {LiBA05c} Li, B.A. \& Chen, L.W. 2005, Phys. Rev. C, 72,
064611

\bibitem[{Li \& Steiner (2006)}]{Lis06}Li, B.A. \& Steiner, A.W. 2006, Phys. Lett. B, 642, 436

\bibitem[{Li {et~al.} (2007)}] {LiBA07}Li, B.A., Chen, L.W., Ma, H.R., Xu, J., \& Yong, G.C. 2007,
Phys. Rev. C, 76, 051601 (R)

\bibitem[{Li {et~al.} (2008)}] {LCK08} Li, B.A., Chen, L.W., \& Ko, C.M. 2008, Phys. Rep., 464,
113

\bibitem[{Link {et~al.} (1999)}] {Lin99}  Link, B., Epstein, R.I., \& Lattimer, J.M. 1999,
Phys. Rev. Lett., 83, 3362

\bibitem[{Lorenz {et~al.}(1993)}] {Lor93} Lorenz, C.P., Ravenhall, D.G., \& Pethick, C.
J. 1993, Phys. Rev. Lett., 70, 379

\bibitem[{Lynch {et~al.} (2009)}] {Lyn09} Lynch, W.G., et al. 2009, arXiv:0901.0412

\bibitem[{Margueron \& Chomaz (2003)}] {Chomaz03} Margueron, J. \& Chomaz, P. 2003,
Phys. Rev. C, 67, 041602 (R)

\bibitem[{Morrison {et~al.} (2004)}] {Morrison} Morrison, I. A., Baumgarte,  T. W., Shapiro,
S. L., \& Pandharipande, V. R. 2004, ApJ, 617, 135

\bibitem[{Mou (2007)}] {Mou07} Moustakidis,  Ch. C. 2007, Phys. Rev. C, 76, 025805

\bibitem[{M\"uller \& Serot (1995)}]{Mul95} M\"uller, H. \& Serot, B.D., 1995,
Phys. Rev. C, 52, 2072

\bibitem[{Negele \& Vautherin (1973)}] {Neg73} Negele, J. W. \& Vautherin, D. 1973, Nucl. Phys. A, 207,
298

\bibitem[{Newton (2009)}]{New09} Newton, William, 2009, private communications

\bibitem[{Oppenheimer \& Volkoff (1939)}] {Oppen} Oppenheimer,
J., \& Volkoff, G. 1939, Phys. Rev., 55, 374

\bibitem[{Oyamatsu (1993)}] {Oya93} Oyamatsu, K. 1993, Nucl. Phys. A, 561, 431

\bibitem[{Oyamatsu \& Iida (2007)}] {Oya07} Oyamatsu, K. \& Iida, K. 2007, Phys. Rev. C, 75
, 015801

\bibitem[{Pethick \& Ravenhall (1995)}] {Pet95a} Pethick, C. J., Ravenhall, D. G. 1995, Ann. Rev. Nucl. Part.
Sci., 45, 429

\bibitem[{Pethick {et~al.} (1995)}] {Pet95b} Pethick, C. J., Ravenhall, D. G., \& Lorenz, C. P. 1995, Nucl.
Phys. A, 584, 675

\bibitem[{Piekraewicz (2007)}]{Piek07} Piekraewicz, J. 2007, Phys. Rev. C, 76, 064310

\bibitem[{Prakash {et~ al.} (1988)}]{Pra88} Prakash, M., Kuo, T. T. S., \& Das Gupta, S. 1988,
Phys. Rev. C, 37, 2253

\bibitem[{Rabhi {et~al.} (2008)}] {Rab08} Rabhi, A., Provid\^{e}ncia, C., \& Da Provid\^{e}ncia, J.,
arXiv:0810.3395v2 [nucl-th]

\bibitem[{Ravenhall {et~al.} (1983)}] {Rav83} Ravenhall, D. G., Pethick, C. J., \& Wilson, J. R. 1983, Phys. Rev.
Lett., 50, 2066

\bibitem[{Ruster {et~al.} (2006)}] {Rus06} Ruster, S. B., Hempel, M., \& Schaffner-Bielich, J. 2006, Phys.
Rev. C, 73, 035804

\bibitem[{Rutledge {et~al.} (2006)}] {Rut06} Rutledge, R. E., et al. 2006, ApJ, 580, 413

\bibitem[{Samyn {et~al.} (2002)}] {Sam02} Samyn, M., et al.
2002, Nucl. Phys. A, 700, 142

\bibitem[{Samyn {et~al.} (2003)}] {Sam03} Samyn, M., Goriely, S. \& Pearson, J.M.
2003, Nucl. Phys. A, 725, 69

\bibitem[{Samyn {et~al.} (2004)}] {Sam04} Samyn, M., et al.
2004, Phys. Rev. C, 70, 044309

\bibitem[{Samyn {et~al.} (2005)}] {Sam05} Samyn, M., Goriely, S. \& Pearson, J.M.
2005, Phys. Rev. C, 72, 044316

\bibitem[{Shetty {et~al.} (2007)}] {She07} Shetty, D., Yennello, S. J., \& Souliotis, G. A. 2007,
Phys. Rev. C, 75, 034602

\bibitem[{Siemens (1983)}]{Sie83}Siemens, P.J. 1983, Nature, 305, 410

\bibitem[{Steiner {et~al.} (2005a)}] {Ste05} Steiner, A.~W., Prakash, M., Lattimer, J.~M., \& Ellis,
P.~J. 2005a, Phys. Rep., 410, 325

\bibitem[{Steiner \& Li (2005b)}] {Ste05b} Steiner, A. W., Li, B.~A. 2005b, Phys. Rev. C, 72,
041601(R)

\bibitem[{Steiner (2006)}] {Ste06} Steiner, A.W. 2006, Phys. Rev. C, 74, 045808

\bibitem[{Steiner (2008)}] {Ste08} Steiner, A.W. 2008, Phys. Rev. C, 77, 035805

\bibitem[{Stone {et~al.} (2003)}] {Stone} Stone, J.R., Miller, J.C., Koncewicz, R.,
Stevenson, P.D., \& Strayer, M.R. 2003, Phys. Rev. C, 68, 034324

\bibitem[{Stone \& Reinhard (2007)}] {Sto07} Stone, J.R. \& Reinhard, P.-G. 2007, Prog. Part.
Nucl. Phys., 58, 587

\bibitem[{Todd-Rutel \& Piekarewicz(2005)}]{Tod05} Todd-Rutel, B.~G. \& Piekarewicz, J. 2005,
Phys. Rev. Lett., 95, 122501

\bibitem[{Tsang {et~al.} (2001)}] {Tsa01} Tsang, M.B., et al. 2001, Phys. Rev. Lett., 86,
5023

\bibitem[{Tsang {et~al.} (2004)}] {Tsa04} Tsang, M.B., et al. 2004, Phys. Rev. Lett., 92, 062701

\bibitem[{Tsang {et~al.} (2008)}] {Tsa08} Tsang, M.B., et al. 2008, arXiv:0811.3107

\bibitem[{van Dalena {et~al.}(2007)}]{Dal07} van Dalena, E.N.E., Fuchs, C., \& Faessler, A. 2007, Eur. Phys. J.
A, 31, 29

\bibitem[{Watanabe (2005)}]{Wat05} Watanabe, G., Maruyama, T., Sato, K.,  Yasuoka, K., \& Ebisuzaki, T.
2005, Phys. Rev. Lett., 94, 031101

\bibitem[{Welke {et~ al.} (1988)}]{Wel88} Welke, G. M., Prakash, M.,  Kuo, T. T. S., Das Gupta, S. \&
Gale, C. 1988, Phys. Rev. C, 38, 2101

\bibitem[{Worley {et~al.} (2008a)}] {Wor08} Worley, A., Krastev, P.G., \& Li, B.A. 2008a,
ApJ, 685, 390

\bibitem[{Worley {et al.} (2008b)}] {Wor09} Worley, A., Krastev, P.G., \& Li, B.A., 2008b, [arXiv:0812.0408].

\bibitem[{Xu {et~al.} (2007a)}] {Xu07a} Xu, J., Chen, L.W., Li, B.A., \& Ma, H.R. 2007a, Phys. Rev. C, 75, 014607

\bibitem[{Xu {et~al.} (2007b)}] {Xu07c} Xu, J., Chen, L.W., Li, B.A., \& Ma, H.R. 2007b, Phys. Lett. B, 650, 348

\bibitem[{Xu {et~al.} (2008a)}] {Xu07b} Xu, J., Chen, L.W., Li, B.A., \& Ma, H.R. 2008a, Phys. Rev. C, 77, 014302

\bibitem[{Xu {et~al.} (2008b)}] {Xu09} Xu, J., Chen, L.W., Li, B.A., \&
Ma, H.R. 2009, Phys. Rev. C, in press; arXiv:0807.4477v1 [nucl-th]

\bibitem[{Zhang \& Chen (2001)}]{Zha01} Zhang, F.S. \& Chen, L.W. 2001, Chinese Phys. Lett., 18, 142

\bibitem[{Zuo {et~al.} (2003)(2006)}] {Zuo03} Zuo, W., et al. 2003, Phys. Rev. C, 69, 064001;
2006, \textit{ibid}. C, 73, 035208

\end{thebibliography}
\end{document}